\documentclass[onecolumn, a4size, 12pt]{IEEEtran}

\usepackage{amsmath,amsfonts,amssymb}
\usepackage{graphicx}
\usepackage{cite}
\usepackage{multirow}
\usepackage{algorithm}
\usepackage{algorithmic}
\usepackage{subfigure}
\usepackage{multirow}
\usepackage{url}
\usepackage{color}
\usepackage{array}
\usepackage[margin=1in]{geometry}
\usepackage{subfig}
\usepackage{tabularx}

\graphicspath{{./figs/}}

\IEEEoverridecommandlockouts

\newtheorem{theorem}{Theorem}

\newtheorem{definition}{Definition}
\newtheorem{lemma}{Lemma}

\newtheorem{proposition}{Proposition}

\newtheorem{remark}{Remark}

\newcommand{\xE}{\mathbb{E}}
\newcommand{\hud}{\underline h}
\newcommand{\hbdud}{\underline {\mathbf h}}
\newcommand{\diag}{\mathrm{diag}}
\newcommand{\Rbdud}{\underline {\mathbf R}}
\newcommand{\tr}{\mathrm{Tr}}
\newcommand{\ttr}{\text{tr}}
\newcommand{\hbdudhat}{\hat{\underline {\mathbf h}}}
\newcommand{\hbdudtd}{{\underline {\mathbf {\tilde h}}}}
\newcommand{\Ebdud}{\underline {\mathbf E}}
\newcommand{\Cbdud}{\underline {\mathbf C}}
\newcommand{\UT}{\mathrm{UT}}

\newcommand{\blkdiag}{\mathrm{blkdiag}}

\begin{document}
\title{Electromagnetic Lens-focusing Antenna Enabled Massive MIMO: Performance Improvement and Cost Reduction}
\author{
Yong~Zeng,~\IEEEmembership{Student~Member,~IEEE,} Rui~Zhang,~\IEEEmembership{Member,~IEEE,} and~Zhi~Ning~Chen,~\IEEEmembership{Fellow,~IEEE}
\thanks{The authors are with the Department of Electrical and Computer Engineering, National University of Singapore.
Email: \{elezeng, elezhang, eleczn\}@nus.edu.sg}
}
\maketitle

\begin{abstract}
Massive multiple-input multiple-output (MIMO) techniques have been recently advanced to tremendously improve the performance of wireless communication networks. 
However, the use of very large antenna arrays at the base stations (BSs) brings new issues, such as the significantly increased hardware and signal processing costs. In order to reap the enormous gain of  massive MIMO and yet reduce its cost to an affordable level, this paper proposes  a novel system design by integrating  an electromagnetic (EM) lens with the large antenna array, termed the \emph{EM-lens enabled MIMO}.  The  EM lens  has the capability of focusing the power  of an incident wave to a small area of the antenna array, while the location of the focal area varies with the angle of arrival (AoA) of the wave. Therefore, in practical scenarios where the arriving signals from geographically separated users have different AoAs, the EM-lens enabled system provides two new benefits, namely \emph{energy focusing} and \emph{spatial interference rejection}. By taking into account the effects of imperfect channel estimation via pilot-assisted training, in this paper we analytically show that the average received signal-to-noise ratio (SNR) in both the single-user and multiuser uplink transmissions  can be strictly improved by the EM-lens enabled system. Furthermore, we demonstrate that the proposed design makes it possible to considerably reduce the hardware and signal processing costs with only slight degradations in performance. To this end, two complexity/cost reduction  schemes are proposed, which are \emph{small-MIMO processing} with parallel receiver filtering applied over subgroups of antennas to reduce the computational complexity, and channel covariance based \emph{antenna selection} to reduce the required number of radio frequency (RF) chains. Numerical results are provided to corroborate our analysis and show the great potential advantages of our proposed EM-lens enabled MIMO system for next generation cellular networks.
\end{abstract}



\section{Introduction}
Multi-antenna or multiple-input multiple-output (MIMO)  systems have been shown to offer great advantages over conventional  single-antenna systems in point-to-point, single-cell multiuser MIMO, as well as multi-cell MIMO transmissions \cite{36,377,130}. Recently, an even more advanced  multi-antenna technique  known as massive MIMO \cite{373,374,375} has been proposed and is becoming increasingly appealing for the next generation (a.k.a. 5G) wireless communication systems. In massive MIMO systems, antenna arrays with a very large number of elements (say, hundreds or even more) are deployed at the base stations (BSs) so that the spectral efficiency in both the downlink and uplink communications can be dramatically enhanced. 
 Furthermore, in the regime where the number of antenna elements, $M$, is much larger than that of the user terminals (UTs), $K$, the channels of different UTs become asymptotically orthogonal under ``favorable'' propagation conditions \cite{373,469}. As a result, the simple matched filter (MF) processing, i.e., maximal ratio transmission (MRT) in the downlink and maximal ratio combining (MRC) in the uplink,  is optimal \cite{373}. Other notable benefits of massive MIMO include, e.g., the reduced transmission power required to achieve a prescribed quality of service (QoS) \cite{459}, the resilience against failures of individual antenna elements, and the possibility to simplify the multiple-access techniques \cite{375}.

Despite of many  promising benefits,  massive MIMO systems are faced with new challenges, which, if not tackled successfully,  could roadblock their widely deployment in practice. Firstly, the use of ultra-large antenna array incurs a high hardware cost, including the cost associated with the radio frequency (RF) elements such as mixers, amplifiers, D/A and A/D converters at each of the  transmit/receive antennas. This, together with the practically limited space available for antenna installation, may ultimately restrict the number of deployable  antennas $M$ to only  a moderately large value, in which case the channel orthogonality between different UTs does not necessarily hold \cite{468}. For such practical scenarios, it has been shown that the low-complexity MF processing performs considerably worse than  regularized zero-forcing (RZF) precoding or minimum mean-square error (MMSE) filtering  \cite{460,471}. 
However, the computational complexity associated with RZF or MMSE in general grows in a cubic order with $M$, which makes the signal processing cost no longer negligible as $M$ increases. 
Another practical issue for massive MIMO systems is  the increased total energy consumption due to the use of a large number of RF chains \cite{465,467}, which can  even negate the power saving with massive MIMO transmissions \cite{459}.

\begin{figure}
\centering
\includegraphics[scale=0.7]{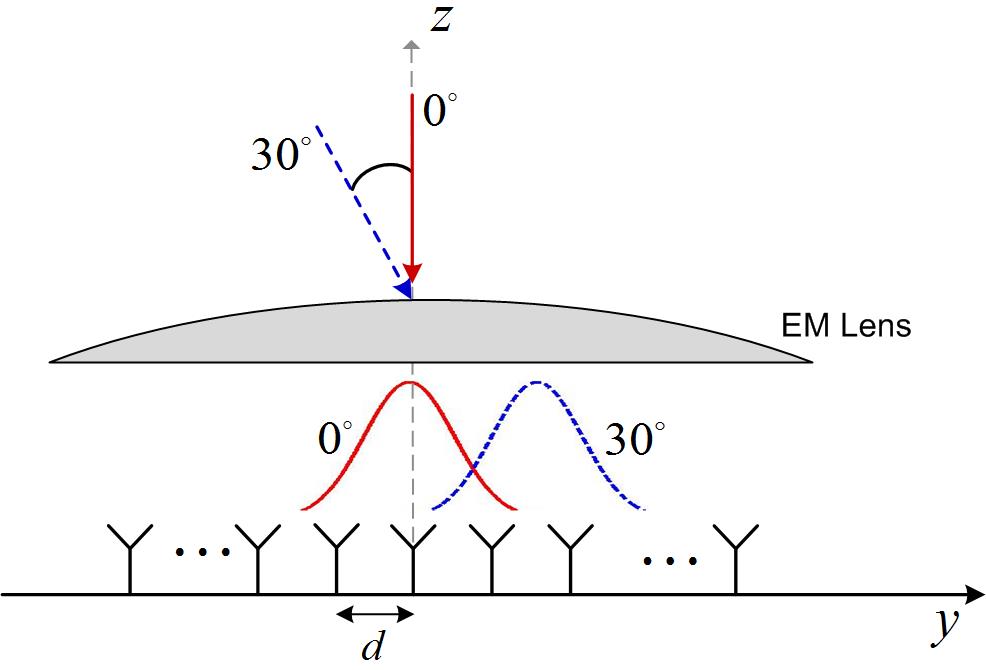}
\caption{Proposed design by integrating EM-lens  with antenna array.}
\label{F:proposed}
\end{figure}

\begin{figure}%
\centering
\includegraphics[scale=0.35]{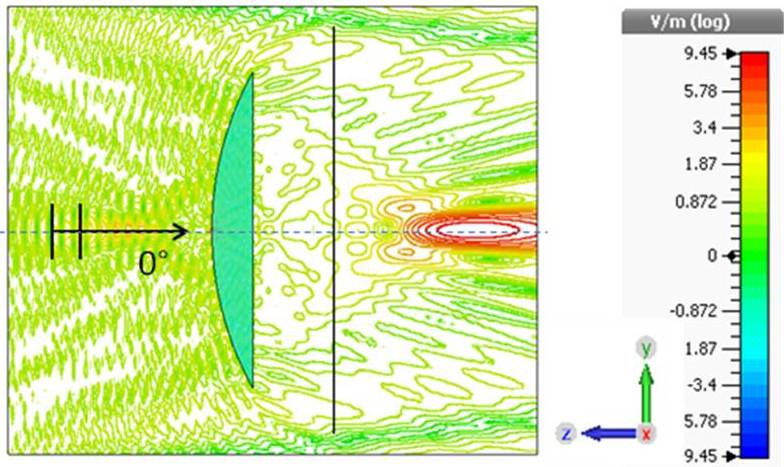}%
\\
(a)
\\
\includegraphics[scale=0.35]{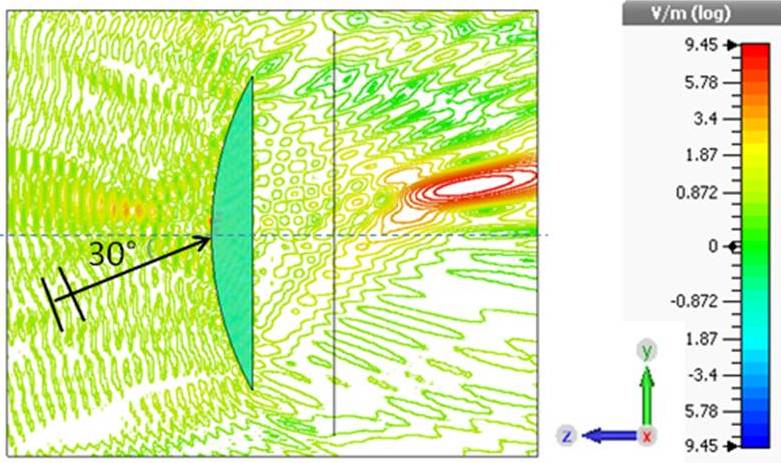}
\\
(b)
\caption{E-field distribution of an EM lens with the AoA of (a) $\theta=0^{\circ}$; and (b) $\theta=30^{\circ}$  \cite{376}.}%
\label{F:RFfileddistribution}%
\end{figure}

In order to capture the promising gains of large MIMO system and yet reduce its cost to an affordable level, we propose in this paper a novel system design by integrating a new component called electromagnetic (EM) lens with the large antenna array, termed {\it EM-lens enabled MIMO}, as shown in Fig.~\ref{F:proposed}.  An EM lens can be practically built using dielectric material with curved front and/or rear surfaces \cite{487,488,376}. With the geometry carefully designed, an EM lens is able to change the paths of incident EM waves in a desired manner so that the arrival signal energy is focused to a smaller region on the antenna array. Furthermore, the spatial power distribution of any incident wave passing through the EM lens  is determined by the angle of arrival (AoA) of the wave. This is demonstrated in  Fig.~\ref{F:RFfileddistribution}, where the E-field distribution of a practical EM lens with  the  refractive index of $2$ is shown \cite{376}. The  aperture diameter and thickness of the EM lens are $12.9 \lambda$ and $1.6\lambda$, respectively, where $\lambda$ is the wavelength in free space. It is observed that as the incident angle $\theta$ changes from $0^{\circ}$ to $30^{\circ}$, the location of the strongest E-field distribution sweeps  accordingly. 
In practice, for the proposed design shown in Fig.~\ref{F:proposed}, the EM lens and the antenna array are integrated and fabricated as a single part, which has the same aperture as the original antenna array but requires extra thickness in order to integrate the EM lens.

In this paper, we apply the proposed EM-lens enabled system to a single-cell multiuser uplink setup under the practical scenario of imperfect channel estimation through uplink training. The performance gain over conventional systems without the EM lens is analytically shown via majorization theory \cite{457}. In particular, for the case of single-user uplink transmission, thanks to {\it energy focusing}, a strictly higher average received signal-to-noise ratio (SNR) is shown to be achievable by the EM-lens enabled system. Moreover, for the general multiuser setup, the performance gain is shown to be twofold: firstly due to energy focusing of the desired user signals as for the single-user case; and secondly due to the {\it spatial interference rejection}, for which the signals of users with sufficiently separated AoAs are effectively discriminated by the AoA-dependent energy focusing of  the EM lens. Furthermore, we demonstrate that the proposed EM-lens enabled system makes it possible to considerably reduce the signal processing and/or hardware costs with only slight degradations in performance. To this end, two complexity/cost reduction schemes are proposed. The first scheme is called {\it small-MIMO processing}, where the receive antennas are divided into groups and the MMSE filtering is performed in parallel over each of the groups with much fewer antenna elements, and hence the total computational complexity is significantly reduced.  In the second scheme, in order to reduce the hardware and energy consumption costs, which in general scale with the number of RF chains each required for one of the active antennas,  we propose a channel covariance based {\it antenna selection}  scheme, with which the number of required RF chains is greatly reduced and excessive training for the conventional instantaneous channel based antenna selection schemes \cite{369} is avoided.

The rest of this paper is organized as follows. Section~\ref{sec:SystemModel} introduces the system model. 
  Section~\ref{sec:channelEstandRate} describes  the channel estimation method and presents the achievable uplink rate with imperfectly estimated channels.  In Section~\ref{sec:performanceAna}, performance analysis based on the average received SNR is given, which shows the advantages of the EM-lens enabled system over the conventional system without the EM lens. Section~\ref{sec:cost} presents two low-complexity/cost techniques, i.e., small-MIMO processing and channel covariance based antenna selection. Numerical  results are given in Section~\ref{sec:simulation}. Finally, we conclude the paper and point out several future working directions in Section~\ref{sec:conclusion}.

\emph{Notations:} $\mathbb{C}^{M\times N}$ and $\mathbb{R}^{M\times N}$  denote the space of $M\times N$ complex and real matrices, respectively. Scalars are denoted by italic letters. Boldface lower- and upper-case letters denote vectors and matrices, respectively. $\mathbf 1$ denotes an all-one vector. $\diag\{\mathbf a\}$ denotes a diagonal matrix with diagonal entries given by vector $\mathbf a$, and $\blkdiag\{\mathbf A_1, \cdots, \mathbf A_n\}$ represents a block diagonal matrix with diagonal blocks given by $\mathbf A_1,\cdots, \mathbf A_n$. $[\mathbf X]_{mn}$ represents the $(m,n)$-th entry
of matrix $\mathbf X$, and $\mathbf X=[x_{mn}]$ denotes a matrix with $(m,n)$-th entries given by $x_{mn}$'s.  For a square matrix $\mathbf{S}$, $\mathrm{Tr}(\mathbf{S})$ denotes its trace, $\boldsymbol \lambda (\mathbf S)$ represents a vector containing all the eigenvalues of $\mathbf S$, and  $\lambda_{\max}(\mathbf S)$ denotes its largest eigenvalue.  For an arbitrary matrix $\mathbf{A}$,  its transpose, Hermitian transpose, and rank are respectively denoted as $\mathbf{A}^{T}$, $\mathbf{A}^{H}$ and $\mathrm{rank}(\mathbf{A})$. $\mathbb{E}[\cdot]$ denotes the expectation operator.  $\mathcal{CN}(\mathbf{x},\mathbf{\Sigma})$ represents the distribution of a circularly symmetric complex Gaussian (CSCG) random vector with mean $\mathbf{x}$ and covariance matrix $\mathbf{\Sigma}$.


\begin{figure}
\centering
\includegraphics[scale=0.65]{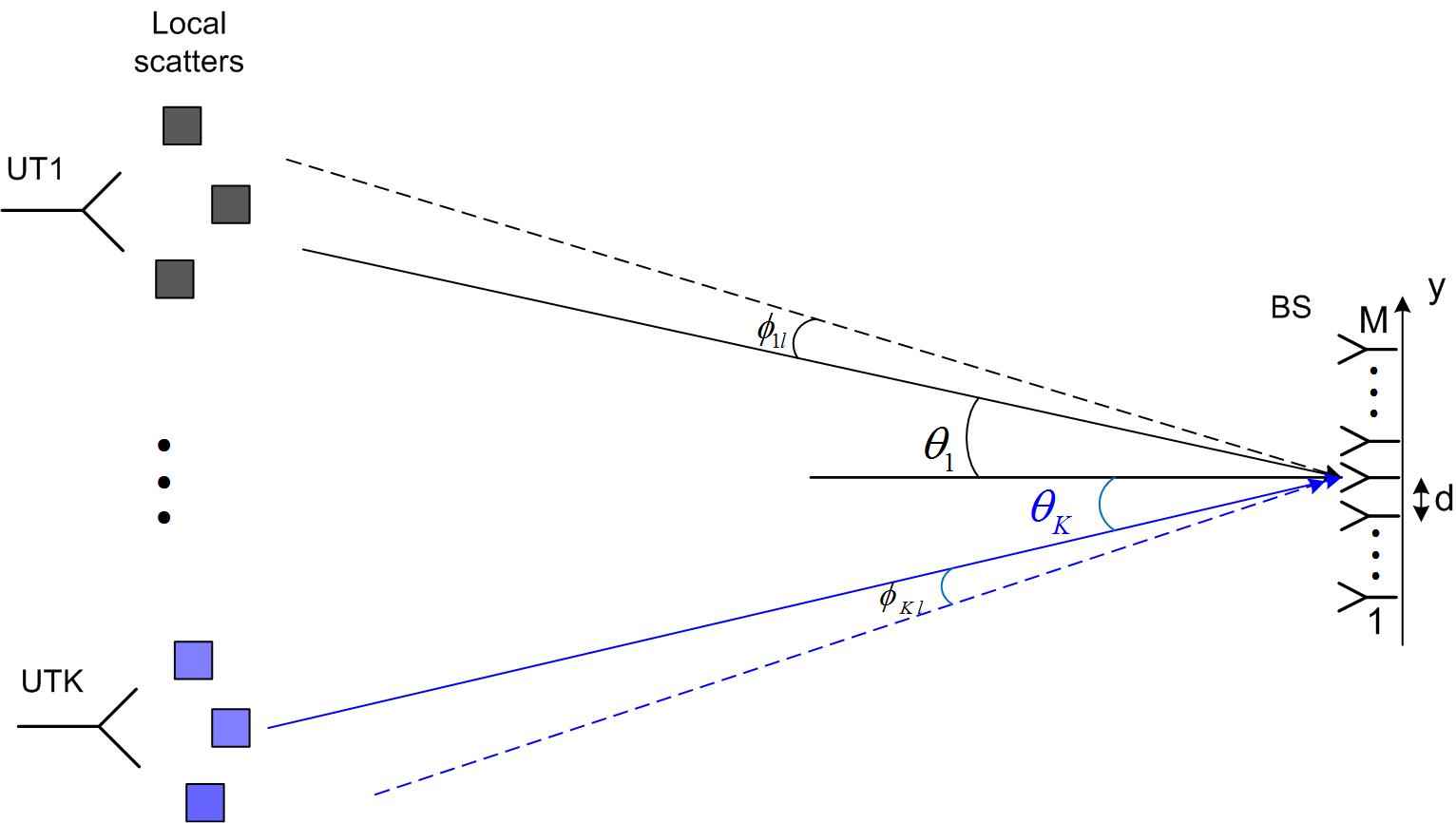}
\caption{Multiuser uplink transmission where the BS is equipped with a uniform linear array.}
\label{F:SingleCellMU}
\end{figure}

\section{System Model}\label{sec:SystemModel}
\subsection{Channel Model without EM Lens}\label{sec:modelnoLens}
First, we consider a single-cell multiuser uplink system as shown in Fig.~\ref{F:SingleCellMU}, where $K$ single-antenna UTs transmit independent messages simultaneously to one BS that is equipped with an $M$-element uniform linear array (ULA). Denote by $d$ the distance between the adjacent elements of the ULA. Without loss of generality, we assume that the ULA is deployed along the y-axis and centered at $y=0$, so that the location $y_m$ of its $m$th element is given by
\begin{align}
y_m=-\frac{(M-1)d}{2}+(m-1)d,\ m=1,\cdots, M.
\end{align}
We assume that the transmitted signal from the $k$th UT ($\UT_k$)  arrives at the BS antenna array via $L_k$ paths, where the $l${th} path, $l=1,\cdots,L_k$, impinges as a plane wave with  AoA $\theta_{kl}$. 
The channel coefficient $h_{km}$ between $\UT_k$ and the $m$th antenna element of the BS can then be expressed as \cite{474}
\begin{align}
h_{km}=\frac{\sqrt{\beta_k}}{\sqrt{L_k}}\sum_{l=1}^{L_k} \sqrt{g_{kl}} \exp\Big(j\frac{2\pi d}{\lambda}{(m-1)\sin\theta_{kl}}\Big), \label{eq:hm}
\end{align}
where $\lambda$ denotes the wavelength, $j$ represents the imaginary unit with $j^2=-1$, $\beta_k$ is the large-scale fading coefficient including the effects of path-loss and shadowing,
 where $\mathbb{E}[|h_{km}|^2]=\beta_k$, $\forall m$,  and $g_{kl}$ is a random variable representing the power gain of the $l$th component for $\UT_k$ with $\frac{1}{L_k}\sum_{l=1}^{L_k} \xE[g_{kl}]=1$, $\forall k$.  We further assume that the AoA $\theta_{kl}$ can be decomposed as $\theta_{kl}=\theta_k+\phi_{kl}$ \cite{440}, where $\theta_k\in [-\Theta, \Theta]$ is the nominal AoA that depends on the location of $\UT_k$, with $\Theta\in (0, \pi]$ denoting the coverage angle of the antenna array,\footnote{For example, in practical cellular systems with sectorized antennas at each BS, we have $\Theta=\pi/3$ for the case of three equally covered sectors in a cell.} and  $\phi_{kl}$ is the AoA offset of the $l$th path relative to $\theta_k$, which is distributed according to a certain power azimuth spectrum (PAS) $f_{\phi}(\phi)$ with zero mean and angular spread (standard deviation) $\sigma_{\phi}$. In practical cellular systems where the BS is elevated in position, $\sigma_{\phi}$ is usually quite small due to the lack of local scatters around the BS. Several distributions have been proposed to approximate the empirically observed PAS, such as the Laplacian  \cite{440} and the Gaussian \cite{458}  distributions.


 Let $\mathbf h_k=\left[\begin{matrix} h_{k1},\cdots,h_{kM}\end{matrix}\right]^T$ denote the channel vector of $\UT_k$ and $\mathbf R_k=\mathbb{E}[\mathbf h_k \mathbf h_k^H]$ be the covariance matrix. Note that since $\mathbb{E}\mathbb[|h_{km}|^2]=\beta_k$, $\forall m$, $\mathbf R_k$ is a positive semidefinite matrix with \emph{identical} diagonal entries equal to $\beta_k$. The $(m,n)$-th entry of $\mathbf R_k$ is given by $[\mathbf R_k]_{mn}=\mathbb E [h_{km}h_{kn}^*]$. 
 As an illustration, if Gaussian  PAS with small $\sigma_{\phi}$ is assumed, closed-form expressions for $[\mathbf R_k]_{mn}$ can be obtained as \cite{458}
 \begin{align}
 [\mathbf R_k]_{mn}=\beta_k & \exp\bigg(-\frac{\sigma_{\phi}^2}{2}\Big(\frac{2\pi d}{\lambda} (m-n)\cos \theta_k\Big)^2\bigg)  \exp\Big(j\frac{2\pi d }{\lambda } (m-n)\sin \theta_k\Big). \label{eq:Gaussian}
 \end{align}

 As $L_k\rightarrow \infty$, by applying the central limit theorem to \eqref{eq:hm}, it follows that $\mathbf h_k$ is zero-mean CSCG distributed with covariance matrix $\mathbf R_k$, i.e.,
 $\mathbf h_k\sim \mathcal{CN}(\mathbf 0, \mathbf R_k)$.

 As can be seen from \eqref{eq:Gaussian}, in the extreme case when $\sigma_{\phi}=0$, which corresponds to the line of sight (LOS) environment, we have $\left|[\mathbf R_k]_{mn}\right|=\beta_k$, $\forall m,n$, i.e., the signals received by different antennas are completely correlated.  On the other hand, when $\sigma_{\phi}\neq 0$, we have $\mathbf R_k\rightarrow \beta_k \mathbf I_M$ as $d\rightarrow \infty$, i.e., independent and identically distributed (i.i.d.) channels are obtained when the antenna elements are sufficiently separated. Therefore, the channel model given in \eqref{eq:hm} with even small angular spread $\sigma_{\phi}$ is still able to include the scenarios ranging from the spatially correlated channels to i.i.d. channels.

  \subsection{Channel Model with EM Lens}
Next, we consider the proposed design where an EM lens is deployed with the ULA at the BS as shown in Fig.~\ref{F:proposed}. In this case, $h_{km}$ given in \eqref{eq:hm} is modified as
\begin{align}\label{eq:hudm}
\hud_{km}=&\frac{\sqrt {\beta_k}}{\sqrt {L_k}} \sum_{l=l}^{L_k} \sqrt{a_{m}(\theta_{kl})}\sqrt{g_{kl}}\exp\Big(j\frac{2\pi d}{\lambda}(m-1) \sin \theta_{kl}\Big),
\end{align}
where the additional factor $a_{m}(\theta_{kl})$ reflects the effect of the AoA-dependent energy focusing by the EM lens, with $a_m(\theta_{kl})/M$ representing the fraction of the power captured by the $m$th antenna element for an incident wave with AoA $\theta_{kl}$. Due to conservation of power, we have $\sum_{m=1}^M a_m(\theta_{kl})=M$, $\forall \theta_{kl}$.
 With small angular spread $\sigma_{\phi}$ for each $\UT_k$, we may apply the approximation $a_m(\theta_{kl})\approx a_m(\theta_k)$, $\forall l$. As a result, \eqref{eq:hudm} can be simplified as
 \begin{equation}  \label{E:hudm2}
\begin{aligned}
\hud_{km}& \approx \sqrt{a_m(\theta_k)}\frac{\sqrt {\beta_k}}{\sqrt {L_k}} \sum_{l=1}^{L_k} \sqrt{g_{kl}}  \exp\Big(j\frac{2\pi d}{\lambda}(m-1) \sin \theta_{kl}\Big)\\
&=\sqrt{a_m(\theta_k)}h_{km}.
\end{aligned}
\end{equation}

\begin{figure}
\centering
\includegraphics[scale=0.75]{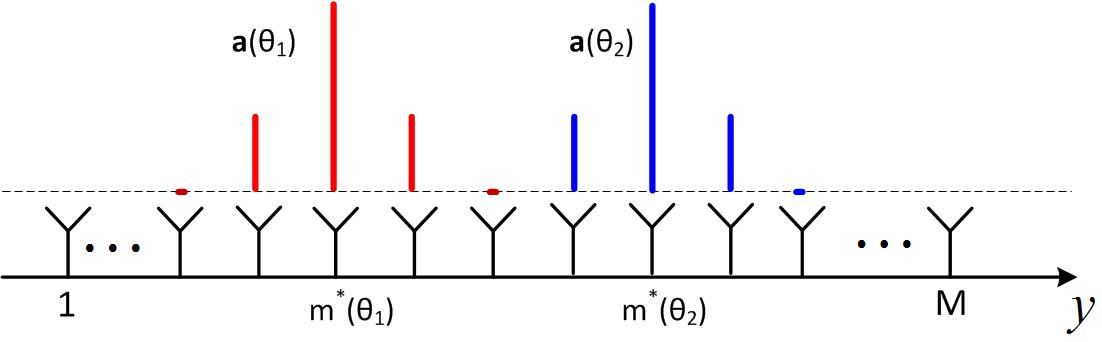}
\caption{An example of the power distribution vectors with $\Delta=1$ for two different AoAs $\theta_1<\theta_2$.}
\label{F:aThetaVector}
\end{figure}

Define the spatial  power distribution vector $\mathbf a(\theta)$ as a function of the AoA $\theta$ as $\mathbf a(\theta)=\left[\begin{matrix}a_1(\theta), \cdots,  a_M(\theta)\end{matrix}\right]^T$, and let $\mathbf A(\theta)=\diag\{\bf a(\theta)\}$. According to \cite{376}, we make some general assumptions in the following for the EM-lens induced power distribution function $\mathbf a(\theta)$, although   most of the results developed later  in this paper hold even without such assumptions.

\emph{Assumption 1:} For any given AoA $\theta\in [-\Theta, \Theta]$, the power distribution vector of the EM-lens enabled system satisfies $\mathbf a(\theta)\neq \mathbf 1$ and furthermore,
\begin{enumerate}
\item $a_m(\theta)\geq a_n(\theta)$, $\forall  |m- m^{\star}(\theta)| \leq |n- m^{\star}(\theta)|$, where $m^{\star}(\theta)= \mathrm{arg}\underset{1\leq m'\leq M}{\max}a_{m'}(\theta)$ denotes the AoA-dependent peak power location that satisfies: $-\Theta\leq \theta_1 \leq \theta_2 \leq \Theta \implies m^{\star}(\theta_1)\leq m^{\star}(\theta_2)$;
\item $a_m(\theta)=0$, $\forall |m-m^{\star}(\theta)|>\Delta$, for some $\Delta$ satisfying $\Delta\leq \min\{m^{\star}(-\Theta)-1, M-m^{\star}(\Theta)\}$.
\end{enumerate}
Assumption 1 is illustrated by Fig.~\ref{F:aThetaVector}, where the power distribution vectors with $\Delta=1$ are shown for two different AoAs $-\Theta<\theta_1<\theta_2<\Theta$. With Assumption 1-(i), we assume that the fraction of the power captured by each antenna element is non-increasing with its distance from the peak power location, which  shifts to the right along y-axis as the AoA increases. With Assumption 1-(ii), we assume that for a given AoA $\theta\in [-\Theta,\Theta]$, the energy is focused to a subset of at most $2\Delta+1$ antennas after passing through the EM lens, where in practice $\Delta$ is a parameter depending on the design of the EM lens and how it is integrated with the antenna array.
Practically,  $\mathbf a(\theta)$ can be modeled by a continuous power density function $f(y;\theta)$ as \cite{456}
\begin{align}\label{eq:atheta}
a_m(\theta)=\begin{cases}
c \int_{y_m-d/2}^{y_m+d/2} f(y;\theta) dy, \ & |m-m^{\star}(\theta)|\leq \Delta,\\
0, & \text{ otherwise},
\end{cases}
\end{align}
where $y_m$ is the location of the $m$th BS antenna and $c$ is a constant such that $\sum_{m=1}^M a_m(\theta)=M$, $\forall \theta$. Based on the results in \cite{376}, for our numerical examples given later in Section~\ref{sec:simulation}, $f(y;\theta)$ is modeled as a Gaussian power density function with mean $\bar y(\theta)$ and variannce $V$, which respectively specify the peak power location and average power spread for an incident wave with AoA $\theta$, i.e.,
\begin{align}\label{eq:fytheta}
f(y;\theta)=\frac{1}{\sqrt{2\pi V}}\exp\Big(-\frac{(y-\bar y(\theta))^2}{2V}\Big).
\end{align}



The channel vector of $\UT_k$ in the EM-lens enabled system is then represented as $\hbdud_k=\left[ \begin{matrix} \hud_{k1},  \cdots , \hud_{kM}\end{matrix}\right]^T=\sqrt{\mathbf A(\theta_k)} \mathbf h_k$, and the covariance matrix of $\hbdud_k$ is given by
\begin{align}\label{eq:Rkudbd}
\Rbdud_k=\xE[\hbdud_k\hbdud_k^H]=\sqrt{\mathbf A(\theta_k)}\mathbf R_k \sqrt{\mathbf A(\theta_k)}.
\end{align}
It is observed from \eqref{eq:Rkudbd} that with the EM lens, the effective channel covariance matrix of each $\UT_k$ is modified based on the power distribution function $\mathbf a(\theta)$ and its AoA $\theta_k$.
As $L_k\rightarrow \infty$, we have $\hbdud_k\sim\mathcal{CN}(\mathbf 0, \Rbdud_k)$. With $\sum_{m=1}^M a_m(\theta_k)=M$ and $[\mathbf R_k]_{mm}=\beta_k$, $\forall m$, it follows that
\begin{align}
\tr(\Rbdud_k)=\tr(\mathbf R_k)= \beta_k M,\label{eq:trR}
\end{align}
which is expected as the EM lens only changes the power distribution of the received signal from
$\UT_k$ on the ULA, while the total signal energy received by the ULA should remain unchanged given the same aperture area of the ULA with or without the EM lens.

Since the original channel vector $\mathbf h_k$ without the EM lens  can be viewed as a special case of $\hbdud_k$ by setting $\mathbf a(\theta_k)=\mathbf 1$, or $\mathbf A(\theta_k)=\mathbf I_M$, in the following sections, we present our results mainly based on the more general channel representation given by $\hbdud_k$.


\section{Uplink Channel Estimation and Achievable Rate}\label{sec:channelEstandRate}
\subsection{Channel Estimation}\label{sec:channelEst}
We assume that the channel covariance matrices $\Rbdud_k$'s are perfectly known at the BS since such second-order channel  statistics vary slowly with time and hence are relatively easy to be estimated in practice. On the other hand, the instantaneous channel vectors $\hbdud_k$'s are estimated at the BS via uplink training. Denote by $\tau$ the number of symbol durations used for training for each coherent block. We assume that orthogonal pilot signals $\mathbf S=\big[\begin{matrix}\mathbf s_1, \cdots, \mathbf s_K\end{matrix}\big]$ with $\mathbf S^H \mathbf S=\mathbf I_K$ are transmitted during the training period, where $\mathbf s_k^H\in \mathbb{C}^{1\times \tau}$ is the pilot sequence transmitted by $\UT_k$. We then have
\begin{align}
\mathbf Y^{\ttr}=\sum_{k=1}^K \sqrt{\rho_{\ttr}} \hbdud_k \mathbf s_k^H + \mathbf N^{\ttr},
\end{align}
where $\mathbf Y^{\ttr}\in \mathbb{C}^{M\times \tau}$ contains the received signals at the BS during the $\tau$ training symbol durations, $\mathbf N^{\ttr}\in \mathbb{C}^{M\times \tau}$ is the additive noise with i.i.d. entries each with normalized power of one, and $\rho_{\ttr}$ represents the training SNR. To estimate the channel for $\UT_k$, the BS projects $\mathbf Y^{\ttr}$ on $\mathbf s_k$ to get a sufficient statistics for estimating $\hbdud_k$. After scaling by $1/\sqrt{\rho_{\ttr}}$, the resulting signal based on which $\hbdud_k$ is estimated can be expressed as
 \begin{align}\label{eq:ytrSC}
 \mathbf y_k^{\ttr}=\frac{1}{\sqrt{\rho_{\ttr}}}\mathbf Y^{\ttr} \mathbf s_k=\hbdud_k +\frac{1}{\sqrt{\rho_{\ttr}}}{\mathbf n_k^{\ttr}},
 \end{align}
 where $\mathbf n_k^{\ttr}\sim \mathcal{CN}(\mathbf 0, \mathbf{I}_M)$.
 The MMSE estimate $\hbdudhat_k$ of $\hbdud_k$ is then given by  \cite{71}
\begin{align}
\hbdudhat_k &=\xE\left[\hbdud_k \mathbf y_k^{\ttr H}\right]\left(\xE\left[\mathbf y_k^{\ttr}\mathbf y_k^{\ttr H}\right]\right)^{-1}{\mathbf y_k^{\ttr}}\\
&=\Rbdud_k\Big(\Rbdud_k+\frac{1}{\rho_{\ttr}}\mathbf I_M\Big)^{-1}{\mathbf y_k^{\ttr}}.\label{eq:hkhat}
\end{align}
Let $\hbdudtd_k$ denote the channel estimation error, i.e., $\hbdudtd_k=\hbdud_k-\hbdudhat_k$. Based on the well-known orthogonal property of the MMSE estimation \cite{71}, we have that $\hbdudtd_k$ and $\hbdudhat_k$ are uncorrelated. Furthermore, since $\hbdud_k$ is CSCG distributed, the distributions of $\hbdudtd_k$ and $\hbdudhat_k$ are respectively given by
\begin{align}
\hbdudtd_k & \sim \mathcal{CN}(\mathbf 0, \Ebdud_k),\\
\hbdudhat_k & \sim \mathcal{CN}(\mathbf 0, \Cbdud_k), \label{eq:hhatDist}
\end{align}
where
\begin{align}
\Ebdud_k&=\xE\left[\hbdudtd_k \hbdudtd_k^H\right]=\Rbdud_k-\Rbdud_k \Big(\Rbdud_k+\frac{1}{\rho_{\ttr}}\mathbf I_M \Big)^{-1} \Rbdud_k, \label{eq:Ematrix}\\
\Cbdud_k&=\xE\left[\hbdudhat_k \hbdudhat_k^H\right]=\Rbdud_k \Big(\Rbdud_k+\frac{1}{\rho_{\ttr}}\mathbf I_M\Big)^{-1} \Rbdud_k \label{eq:Cmatrix}.
\end{align}

\subsection{Achievable Rate}
After the training based channel estimation, uplink data transmission from the UTs follows. The signal received at the BS can be expressed as
\begin{align}\label{eq:y}
\mathbf y=\sqrt{\rho_{d}} \hbdud_k x_k + \sum_{u\neq k} \sqrt{\rho_{d}} \hbdud_u x_u +\mathbf n,
\end{align}
where $\rho_{d}$ denotes the SNR for the uplink data communication, $x_k$ is the information symbol from $\UT_k$ with normalized power of one, and
 $\mathbf n\sim \mathcal{CN}(\mathbf 0, \mathbf I_M)$ represents the additive noise.
Let $\mathbf v_k\in \mathbb{C}^{M\times 1}$ denote the linear filter applied at the BS for detecting the signal transmitted from $\UT_k$. We then have
\begin{equation}\label{eq:xkhat}
\begin{aligned}
\hat x_k =& \mathbf v_k^H \mathbf y \\
=&\sqrt{\rho_d} \mathbf v_k^H \hbdudhat_k x_k + \sqrt{\rho_d} \sum_{u\neq k}\mathbf v_k^H \hbdudhat_u x_u + \sqrt{\rho_d} \sum_{u=1}^K \mathbf v_k^H \hbdudtd_u x_u+  \mathbf v_k^H \mathbf n,
\end{aligned}
\end{equation}
where we have used the identity $\hbdud_k=\hbdudhat_k+\hbdudtd_k$, $\forall k$. Since the BS only knows  the estimated channel vectors $\{\hbdudhat_k\}_{k=1}^K$, only the first term in \eqref{eq:xkhat} is treated as the desired signal from  $\UT_k$, and all the remaining terms, which are uncorrelated with the desired signal term, are treated as noise \cite{460,463}.  
 Following the standard bounding technique based on the worst-case uncorrelated noise \cite{2}, the uplink achievable rate for $\UT_k$ is given by
\begin{align}\label{eq:Rk}
R_k=\mathbb{E}\left[\log_2\left(1+\gamma_k\right)\right],
\end{align}
where the received SNR $\gamma_k$ for a given channel realization is
\begin{align}
\gamma_k=\frac{|\mathbf v_k^H \hbdudhat_k |^2}{\mathbf v_k^H \left( \sum_{u\neq k} \hbdudhat_u \hbdudhat_u^H + \sum_{u=1}^K \Ebdud_u+\frac{1}{\rho_{d}}\mathbf I_M\right) \mathbf v_k}.\label{eq:gamma}
\end{align}
From \eqref{eq:gamma}, the optimal $\mathbf v_k$ that maximizes  $\gamma_k$ is the MMSE filter given by
\begin{align}\label{eq:vkMMSE}
\mathbf v_k=\Big( \sum_{u\neq k} \hbdudhat_u \hbdudhat_u^H + \sum_{u=1}^K \Ebdud_u+\frac{1}{\rho_{d}}\mathbf I_M\Big)^{-1} \hbdudhat_k,
\end{align}
and the corresponding maximum SNR is
\begin{align}
\gamma_k=\hbdudhat_k^H \Big( \sum_{u\neq k} \hbdudhat_u \hbdudhat_u^H + \sum_{u=1}^K \Ebdud_u+\frac{1}{\rho_{d}}\mathbf I_M\Big)^{-1} \hbdudhat_k. \label{eq:gammaMMSE}
\end{align}

\section{Performance Analysis}\label{sec:performanceAna}
As the achievable rate $R_k$ given in \eqref{eq:Rk} is difficult to characterize for finite system dimensions, most existing analytical results in the literature are based on the deterministic approximations of the SNR by assuming that $M$ and $K$ both grow infinitely large while keeping a fixed ratio $K/M$ \cite{460}. An alternative approach that  works for finite system dimensions is based on the average received SNR \cite{464}. In this section, by adopting the average received SNR as our performance metric, we compare the performance for the MIMO systems with versus without the EM lens.

Note that the received SNR $\gamma_k$ given in \eqref{eq:gammaMMSE} varies with the estimated channel vectors $\{\hbdudhat_u\}_{u=1}^K$, which in turn depend on the channel realizations $\{\hbdud_u\}_{u=1}^K$. The average received SNR, $\mathbb{E} [\gamma_k]$, where the expectation is taken over the channel realizations, is then given by
\begin{align}
\mathbb{E}[\gamma_k] 
&=\mathbb{E}  \bigg[\tr\bigg(\Big( \sum_{u\neq k} \hbdudhat_u \hbdudhat_u^H + \sum_{u=1}^K \Ebdud_u+\frac{1}{\rho_{d}}\mathbf I_M\Big)^{-1} \hbdudhat_k\hbdudhat_k^H \bigg)\bigg]\\
&=\tr  \bigg(\mathbb{E} \Big[ \Big(\sum_{u\neq k} \hbdudhat_u \hbdudhat_u^H + \sum_{u=1}^K \Ebdud_u+\frac{1}{\rho_{d}}\mathbf I_M\Big)^{-1}\Big] \mathbb{E}[\hbdudhat_k \hbdudhat_k^H]\bigg) \label{eq:indhkhu} \\
&\geq \tr  \bigg(\Big( \sum_{u\neq k} \Cbdud_u + \sum_{u=1}^K \Ebdud_u+\frac{1}{\rho_{d}}\mathbf I_M\Big)^{-1} \Cbdud_k \bigg)  
 \triangleq \bar{\gamma}_k,\label{eq:gammaLB}
\end{align}
where \eqref{eq:indhkhu} is due to the commutativity between the two operators $\mathbb{E}[\cdot]$ and $\tr(\cdot)$, as well as the independence between $\hbdudhat_u$ and $\hbdudhat_k$ for $u\neq k$; \eqref{eq:gammaLB} follows from the Jensen's inequality and the fact that $\tr(\mathbf X^{-1}\Cbdud_k)$ is a convex function with respect to any  positive definite matrix $\mathbf X$ \cite{202}. 
In general, $\bar{\gamma}_k$ defined in \eqref{eq:gammaLB} is a lower bound for the average received SNR $\mathbb{E}[\gamma_k]$, since it is obtained by discarding the estimated channel knowledge of all other UTs when detecting the signal for $\UT_k$. In this case, each of the terms $\hbdudhat_u\hbdudhat_u^H$, $u\neq k$, in \eqref{eq:gamma} and \eqref{eq:vkMMSE} is replaced by its statistical expectation $\Cbdud_u$.
Since both $\Ebdud_u$ and $\Cbdud_u$ are related to the channel covariance matrix $\Rbdud_u$ via \eqref{eq:Ematrix} and \eqref{eq:Cmatrix}, respectively, $\bar{\gamma}_k$ is a function of the $K$ covariance matrices $\{\Rbdud_u\}_{u=1}^K$, and hence is explicitly denoted as $\bar \gamma_k\left(\Rbdud_1,\cdots, \Rbdud_K\right)$. 
In the following two subsections, we analytically show the performance gain of the EM-lens enabled system based on $\bar \gamma_k\left(\Rbdud_1,\cdots, \Rbdud_K\right)$, first for the single-user case, and then for the more general multiuser setup.

\subsection{Single-User System}
First, consider the single-user setup  with $K=1$. In this case, no inter-user interference is present, and thus the inequality in \eqref{eq:gammaLB} becomes equality and $\bar \gamma_k$ is exactly equal to the average received SNR, i.e.,
\begin{align}
\mathbb{E}[\gamma]=\bar{\gamma}(\Rbdud)=\tr  \Big(\big(\Ebdud +\frac{1}{\rho_d} \mathbf I_M\big)^{-1} \Cbdud \Big).\label{eq:gammaSU}
\end{align}
Note that for brevity, we have dropped the user index $k$ since $k=1$ in this subsection.

\begin{lemma}\label{lemma:SchurCVX}
For the single-user system, the average received SNR $\mathbb{E}[\gamma]=\bar{\gamma} (\Rbdud)$ depends on the channel covariance matrix $\Rbdud$ only through its eigenvalues, i.e., $\bar{\gamma}(\Rbdud)=f\left(\boldsymbol \lambda(\Rbdud)\right)$ for some  function $f: \mathbb{R}_+^M\rightarrow \mathbb R$. Furthermore,  $f(\mathbf x)$ is given by
\begin{align}
f(\mathbf x)=\sum_{m=1}^M \frac{\rho_d \rho_{\ttr} x_m^2}{\left(\rho_d+\rho_{\ttr}\right)x_m+1},\label{eq:fx}
\end{align}
which is a strictly Schur-convex function for finite  $\rho_{\ttr}$ and $\rho_d$.\footnote{For a brief introduction of Schur-convex function and the associated majorization theory, please refer to Appendix~\ref{A:major}.}
\end{lemma}
\begin{IEEEproof}
With \eqref{eq:Ematrix} and \eqref{eq:Cmatrix}, we have $\Ebdud=\Rbdud-\Cbdud$, which yields
\begin{align}
 \Big(\Ebdud+\frac{1}{\rho_d}\mathbf I_M\Big)^{-1}\Cbdud
&=\Big(\Rbdud+\frac{1}{\rho_d}\mathbf I_M-\Cbdud\Big)^{-1}\Cbdud\\
&=\bigg(\mathbf I_M-\Big(\Rbdud+\frac{1}{\rho_d}\mathbf I_M\Big)^{-1}\Cbdud\bigg)^{-1}-\mathbf I_M \label{eq:matrixInvLemma}\\
&=\mathbf U \left[\bigg(\mathbf I_M- \Big(\mathbf \Lambda+\frac{1}{\rho_d}\mathbf I_M\Big)^{-1}\mathbf \Lambda ^2 \Big(\mathbf \Lambda+\frac{1}{\rho_{\ttr}}\mathbf I_M\Big)^{-1} \bigg)^{-1}-\mathbf I_M \right] \mathbf U^H, \label{eq:subEVD}
\end{align}
where \eqref{eq:matrixInvLemma} is due to the matrix inversion lemma \cite{361}, and \eqref{eq:subEVD} is obtained by substituting the identity \eqref{eq:Cmatrix}, as well as the eigenvalue decomposition $\Rbdud=\mathbf U \mathbf \Lambda \mathbf U^H$. By substituting \eqref{eq:subEVD} into \eqref{eq:gammaSU} and with the identity $\tr (\mathbf X)=\sum_{m=1}^M \lambda_m\left(\mathbf X\right)$, we get
\begin{align}
\bar \gamma(\Rbdud) &=\sum_{m=1}^M \frac{\rho_d \rho_{\ttr} \lambda_m^2(\Rbdud)}{\left(\rho_d+\rho_{\ttr}\right)\lambda_m(\Rbdud)+1}.\label{eq:lambdaR}
\end{align}
 Therefore, \eqref{eq:fx} follows. Furthermore, the strict Schur-convexity of $f(\mathbf x)$ can be easily verified using Lemma~\ref{lemma:schurcvxTest} presented in Appendix~\ref{A:major}.

 This thus completes the proof of Lemma~\ref{lemma:SchurCVX}.
\end{IEEEproof}

With Lemma~\ref{lemma:SchurCVX}, we immediately have the following result.
\begin{theorem}\label{theo:strictImp}
For the single-user system with finite  $\rho_{\ttr}$ and $\rho_d$, we have
$\bar{\gamma}(\mathbf R)< \bar{\gamma}(\Rbdud)$
if $\boldsymbol \lambda(\mathbf R) \prec \boldsymbol \lambda(\Rbdud)$ and $\boldsymbol \lambda(\Rbdud)$ is not a permutation of $\boldsymbol \lambda(\mathbf R)$.
\end{theorem}

Theorem~\ref{theo:strictImp} states that a strict performance gain in terms of the average received SNR is achieved by the EM-lens enabled system if the power distribution vector $\mathbf a$ can be designed so that the eigenvalues of the new channel covariance matrix $\Rbdud=\diag\{\sqrt{\mathbf a}\}\mathbf R \diag\{\sqrt{\mathbf a}\}$ majorizes those of $\mathbf R$. This in general holds for practical power distribution functions of $\mathbf a$ (e.g., those satisfying Assumption 1), thanks to the energy focusing property of the EM-lens enabled system. In the following, we compare $\bar \gamma(\Rbdud)$ for the EM-lens enabled system with $\bar \gamma(\mathbf R)$ for the conventional system without EM lens under various conditions, in order to draw more insights to the result in Theorem~\ref{theo:strictImp}.

\begin{proposition}\label{prop:rhoInfty}
For $\rho_{\ttr}\rightarrow \infty$ or $\rho_d \rightarrow \infty$, we have $\bar{\gamma}(\mathbf R)= \bar{\gamma}(\Rbdud)$, $\forall \mathbf a$.
\end{proposition}
\begin{IEEEproof}
It can be easily obtained from Lemma~\ref{lemma:SchurCVX} that as $\rho_{\ttr}\rightarrow \infty$ (or $\rho_d\rightarrow \infty$),  we have $\bar \gamma(\Rbdud)=\rho_d\sum_{m=1}^M \lambda_m(\Rbdud)=\rho_d\tr(\Rbdud)=\rho_d \beta M$ (or $\rho_{\ttr} \beta M$), regardless of the power distribution vector $\mathbf a$. The proof is thus completed.
\end{IEEEproof}

Proposition~\ref{prop:rhoInfty} shows that no performance gain is achieved by the EM-lens enabled system with infinite power for training or data transmission, which is quite intuitive since energy focusing by the EM lens provides no benefit when unlimited power is available.

\begin{proposition}\label{prop:LOS}
For the single-user system with LOS channel, i.e., $|[\mathbf R]_{mn}|=\beta$, $\forall m,n=1,\cdots, M$, we have $\bar \gamma(\mathbf R)=\bar\gamma(\Rbdud)$, $\forall \mathbf a$.
\end{proposition}
\begin{IEEEproof}
Please refer to Appendix~\ref{A:LOS}.
\end{IEEEproof}

Proposition~\ref{prop:LOS} states that no performance gain is achieved by the EM-lens enabled system in LOS environment, since in this case the receive antennas are completely correlated and hence energy focusing to one particular subset of antennas provides no gain in average received SNR.

\begin{proposition}\label{prop:optimal}
For the single-user system with non-LOS channel, i.e., $\mathrm{rank}(\mathbf R)>1$, we have $\bar{\gamma}(\mathbf R)< \bar{\gamma}(\Rbdud)$ if $\mathbf a^{\star}=M\mathbf e_m$ for some $m\in \{1,\cdots, M\}$, where $\mathbf e_m$ is the $m$th column of $\mathbf I_M$.  In this case, the average received SNR is given by
\begin{align}
\bar \gamma^{\star}= \frac{ \rho_d \rho_{\ttr} \beta^2 M^2}{\left(\rho_d+\rho_{\ttr}\right)\beta M+1}. \label{eq:gammaOpt}
\end{align}
\end{proposition}
\begin{IEEEproof}
With \eqref{eq:trR}, we have that the sum of the eigenvalues of $\mathbf R$ is $\mathbf 1^T \boldsymbol \lambda(\mathbf R)=\tr(\mathbf R)=\beta M$. Based on Lemma~\ref{lemma:majortrivial} in Appendix~\ref{A:major}, the following majorization  relation holds:
\begin{equation}
\begin{aligned}\label{eq:lambdasIdealFocus}
\boldsymbol \lambda(\mathbf R)\prec & \left[\begin{matrix} \beta M & 0 & \cdots & 0 \end{matrix} \right]^T =\boldsymbol \lambda(\diag\{\sqrt{M} \mathbf e_m \}\mathbf R \diag\{\sqrt{M} \mathbf e_m \}),
\end{aligned}
\end{equation}
where the right hand side of \eqref{eq:lambdasIdealFocus} is not a permutation of $\boldsymbol \lambda(\mathbf R)$ if and only if $\mathrm{rank} (\mathbf R)>1$. Together with Theorem~\ref{theo:strictImp}, Proposition~\ref{prop:optimal} thus follows.
\end{IEEEproof}

Proposition~\ref{prop:optimal} affirms that the EM-lens enabled system yields a strict performance gain if the channel is non-LOS and moreover, the power distribution vector $\mathbf a^{\star}$ corresponds to an ``ideal'' EM lens, i.e., all energy of the signal passing through the lens  is focused on one single antenna.

\begin{remark}
  For the conventional massive MIMO system without EM lens,  it has been shown in \cite{459} that when the number of BS antennas $M$ grows to infinity, the transmit power of the UT can be asymptotically reduced proportionally to $1/M$ if the BS has perfect CSI, and proportionally to $1/\sqrt{M}$ if CSI is estimated from uplink pilots. To revise this result in the EM-lens enabled system,
we let $\rho_{\ttr}=\rho_d=\frac{E_u}{M}$ in \eqref{eq:gammaOpt}, where $E_u$ is a fixed power. We then have
\begin{align}
\bar \gamma^{\star}=\frac{\beta^2 E_u^2}{2\beta E_u+1},
\end{align}
which is a constant not related to $M$. This implies that for the EM-lens enabled system with ideal energy focusing as given in Proposition~\ref{prop:optimal}, the transmit power of the UT can be reduced proportionally to $1/M$ as $M$ increases, without incurring any loss in average received SNR, even with imperfect CSI at the BS. This is in sharp contrast to that obtained for conventional systems without the EM lens in \cite{459}.
 \end{remark}

In practice, the condition of the power distribution function given in Proposition~\ref{prop:optimal} may not be exactly  met due to non-ideal energy focusing. It is therefore of practical interest to study whether a strict performance gain is still achievable by the EM-lens enabled system with less stringent requirement on the energy focusing of the EM lens.




\begin{proposition}\label{prop:iid}
For the single-user system with spatially uncorrelated channel, i.e., $\mathbf R=\beta \mathbf I_M$ and hence $\Rbdud=\beta \mathrm{diag}\{\mathbf a\}$, we have $\bar \gamma(\mathbf R)< \bar \gamma(\Rbdud), \  \forall \mathbf a\neq \mathbf 1$.
\end{proposition}
\begin{IEEEproof}
With $\mathbf R=\beta \mathbf I_M$ and $\Rbdud=\beta \mathrm{diag}\{\mathbf a\}$, by applying Lemma~\ref{lemma:majortrivial} in Appendix~\ref{A:major}, we have
\begin{align}
\boldsymbol \lambda(\mathbf R)=\beta \mathbf 1 \prec \beta \mathbf a = \boldsymbol \lambda(\Rbdud).
\end{align}
Furthermore, $\beta \mathbf a$ is not a permutation of $\beta \mathbf 1$ whenever $\mathbf a\neq 1$. Therefore, Proposition~\ref{prop:iid} immediately follows from Theorem~\ref{theo:strictImp}.
\end{IEEEproof}

Proposition~\ref{prop:iid} states that for spatially uncorrelated channels, energy focusing is always beneficial since in this case, power is unevenly distributed over the receive antenna elements for both channel estimation and data transmission, which helps improve the average received SNR due to its Schur-convexity.


\begin{proposition}\label{prop:lowSNR2}
For the single-user system with non-LOS channel and $\rho_d+\rho_{\ttr}\ll \frac{1}{\beta M}$,  we have $\bar \gamma(\Rbdud)> \bar \gamma(\mathbf R)$ for power distribution vector $\mathbf a$ satisfying Assumption 1.
\end{proposition}
\begin{IEEEproof}
Please refer to Appendix~\ref{A:lowSNR2}.
\end{IEEEproof}

Proposition~\ref{prop:lowSNR2} shows that for the non-LOS channel under the ``low-SNR'' regime, the EM-lens enabled system is more beneficial even with practical EM lens and under spatially correlated channels.

\subsection{Multiuser System}
For multiuser systems with $K>1$, we focus on the case of spatially uncorrelated channels for our analysis, i.e., $\mathbf R_k=\beta_k \mathbf I_M$ and $\Rbdud_k=\beta_k \diag\{\mathbf a(\theta_k)\}$, $\forall k$, while the performance comparison under the more general correlated channels will be shown by simulation results in Section~\ref{sec:simulation}. With  \eqref{eq:Ematrix}, \eqref{eq:Cmatrix} and \eqref{eq:gammaLB}, the lower bound on the average received SNR of $\UT_k$ for uncorrelated channels can be obtained as
\begin{align}
\bar{\gamma}_k\left(\Rbdud_1,\cdots, \Rbdud_K\right)
=\sum_{m=1}^M\frac{\rho_{\ttr}\rho_d \beta_k^2 a_m^2(\theta_k)}{\beta_k a_m(\theta_k)\Big(\rho_{\ttr}\rho_d \sum_{u\neq k} \beta_u a_m(\theta_u)+\rho_{\ttr}+\rho_d \Big)+\rho_d\sum_{u\neq k} \beta_ua_m(\theta_u)+1}.\label{eq:gammaMU}
\end{align}

\begin{theorem}\label{theo:gainMU}
For the multiuser system with spatially uncorrelated channels, i.e., $\mathbf R_k=\beta_k \mathbf I_M$ and hence $\Rbdud_k=\beta_k \diag\{\mathbf a(\theta_k)\}$, $\forall k$, we have $\bar{\gamma}_k\left(\Rbdud_1,\cdots, \Rbdud_K\right)>\bar{\gamma}_k\left(\mathbf R_1,\cdots, \mathbf R_K\right)$ if  the power distribution vectors $\big\{\mathbf a(\theta_k)\big\}_{k=1}^K$ satisfy $\mathbf a(\theta_k)\neq \mathbf 1$, $\forall k$, and
\begin{equation}\label{eq:condMU}
\begin{aligned}
&\Big(a_m(\theta_k)-a_n(\theta_k)\Big)\Big(\sum_{u\neq k} \beta_u a_m(\theta_u)-\sum_{u\neq k} \beta_u a_n(\theta_u)\Big) \leq 0, \ m, n\in \{1,\cdots, M\}.
\end{aligned}
\end{equation}
\end{theorem}
\begin{IEEEproof}
Please refer to Appendix~\ref{A:gainMU}.
\end{IEEEproof}


 An intuitive explanation of Theorem~\ref{theo:gainMU} is as follows. In the conventional system without EM lens, on average, the  power received from each UT is evenly distributed across all the antennas; hence, for each $\UT_k$, the desired signals at all the $M$ receive antennas are equally corrupted by the interference signals from all other UTs. In contrast, with the AoA-dependent energy focusing provided by the EM lens, the received signals from different UTs are focused at different subsets of the receive antennas. Theorem~\ref{theo:gainMU} thus affirms  that a strict performance gain is achievable for each $\UT_k$ if the antenna element with higher (lower) desired signal power, e.g., $a_m(\theta_k)\geq a_n(\theta_k)$, is corrupted by a lower (higher) total interference, i.e., $\sum_{u\neq k} \beta_u a_m(\theta_u)\leq \sum_{u\neq k} \beta_u a_n(\theta_u)$. It can be verified that the conditions in \eqref{eq:condMU} are satisfied for all UTs if their AoAs are sufficiently separated such that for each $\UT_k$, the interference and desired signals have non-overlapping support sets across the antenna array, i.e., $\forall u\neq k$, we have $a_m(\theta_k)a_m(\theta_u)=0, \forall m$. Under Assumption~1, such an ideal spatial interference rejection is achieved if $\underset{u\neq k}{\min} \left|m^{\star}(\theta_k)-m^{\star}(\theta_u)\right|\geq 2\Delta+1$, $\forall k$.

\begin{remark}
Based on the proof of Theorem~\ref{theo:gainMU} given in Appendix~\ref{A:gainMU}, the performance gain of the EM-lens enabled multiuser system is due to two main factors: the energy focusing of desired signals as reflected by the first inequality in \eqref{eq:twoIneq}, and the spatial separation of interfering signals as reflected by the second inequality in \eqref{eq:twoIneq}. In contrast, the performance gain in the single-user system is attributed to the energy focusing of the desired signal only.
\end{remark}

%

\begin{remark}\label{remark:MURhoInfty}
Note that due to spatial interference rejection, Theorem~\ref{theo:gainMU} holds even with infinite transmit power, i.e., $\rho_{\ttr}\rightarrow \infty$ or $\rho_d\rightarrow \infty$. 
This is in contrast to the single-user scenario where the performance gain, which is due to the energy focusing of the desired signals only, vanishes as $\rho_{\ttr}\rightarrow \infty$ or $\rho_d\rightarrow \infty$ (see Proposition~\ref{prop:rhoInfty}).


%
%
\end{remark}

\section{Low-Complexity Design}\label{sec:cost}
In this section, we present two complexity/cost reduction techniques that work particularly well for the proposed EM-lens enabled system. The first one is called small-MIMO processing, which aims to reduce the signal processing complexity of the full-scale MMSE receiver given in \eqref{eq:vkMMSE}. The second one is termed channel covariance based antenna selection, which is designed to reduce the number of required RF chains, and hence saves the hardware and energy consumption costs.

\subsection{Small-MIMO Processing}\label{sec:smallMIMO}
The MMSE receiver given in \eqref{eq:vkMMSE} requires performing an $M$-dimensional matrix inversion, which may incur a high computational complexity for large $M$. In this subsection, we propose a low-complexity multiuser receiver design called small-MIMO processing for the EM-lens enabled system, which is able to considerably reduce the computational complexity as compared to the full-scale MMSE receiver in \eqref{eq:vkMMSE}. 

 For the proposed scheme, the $M$ receive antennas at the BS are divided into $G$ groups, where the $g$th group, $g=1,\cdots, G$, has $M_g$ antennas with $\sum_{g=1}^GM_g=M$. Since the incident power from each UT is focused on a subset of adjacent antennas in the EM-lens enabled system (cf. Assumption 1), we simply assign  adjacent antenna elements with appropriate sizes into different groups so that the signals received from all the antennas are first processed in parallel groups, and then linearly combined over the outputs of different groups, to reduce the overall signal processing complexity. Let the indices of antennas in the $g$th group be denoted by the set $\mathcal{M}_g=\left\{\sum_{g'=1}^{g-1} M_{g'}+1, \sum_{g'=1}^{g-1} M_{g'}+2, \cdots, \sum_{g'=1}^gM_{g'}\right\}$. The received signal vector given in \eqref{eq:y} can then be re-expressed as
\begin{align}
\mathbf y^g=\sqrt{\rho_{d}} \hbdud_k^g x_k + \sum_{u\neq k} \sqrt{\rho_{d}} \hbdud_u^g x_u +\mathbf n^g,\quad  g=1,\cdots,G,
\end{align}
where $\mathbf y^g$, $\hbdud_u^g$, $\mathbf n^g\in \mathbb{C}^{M_g\times 1}$ denote the received signal, channel, and noise vectors corresponding to the antennas in $\mathcal{M}_g$, respectively. By small-MIMO processing, the BS first performs MMSE filtering separately over the received signals within each of the $G$ groups in parallel, and then linearly combines the filtered signals from the $G$ groups. Let $\mathbf v_{k}^g\in \mathbb{C}^{M_g\times 1}$ denote the MMSE filter applied by the $g$th group for detecting the signal from $\UT_k$. Similar to \eqref{eq:vkMMSE}, we have
\begin{align}
\mathbf v_k^g=\mathbf J_k^g \hbdudhat_k^g, \ g=1,\cdots, G, \label{eq:vkg}
\end{align}
where $\mathbf J_k^g=\big(\sum_{u\neq k} \hbdudhat_u^g (\hbdudhat_u^{g})^H + \sum_{u=1}^K \Ebdud_u^g+\frac{1}{\rho_{d}}\mathbf I_{M_g}\big)^{-1}$, $\hbdudhat_u^g$ and $\Ebdud_u^g$ have similar definitions as in Section~\ref{sec:channelEst}, but apply only to  antennas in the $g$th group.  The filtered output for $\UT_k$ from the $g$th antenna group  can be written as
 \begin{align}\label{eq:ykg}
  y_k^g=(\mathbf v_k^g)^H \mathbf y^g = \sqrt{\rho_d} (\mathbf v_k^g)^H \hbdudhat_k^g x_k + I_k^g, \ g=1,\cdots, G,
 \end{align}
  where $I_k^g\triangleq \sqrt{\rho_d} (\mathbf v_k^g)^H \hbdudtd_k^g x_k+\sum_{u\neq k} \sqrt{\rho_d} (\mathbf v_k^g)^H \hbdud_u^g x_u + (\mathbf v_k^g)^H\mathbf n^g$ is the interference-plus-noise term for $\UT_k$ resulting from the $g$th antenna group, with $\hbdudtd_k^g$ denoting the channel estimation error for $\UT_k$ corresponding to the BS antennas in the $g$th group. To detect $x_k$, the filtered outputs in \eqref{eq:ykg} from all the $G$ groups are linearly combined, which gives
\begin{align}\label{eq:xkhatsmallMIMO}
\hat x_k=\sum_{g=1}^G w_k^g y_k^g=\sum_{g=1}^G w_k^g (\hbdudhat_k^{g})^H\mathbf J_k^g \mathbf y^g,
\end{align}
where $w_k^g$ is the combining weight for group $g$. As evident from \eqref{eq:ykg}, with the simple MRC scheme, $w_k^g$ is given by
\begin{align}
 w_k^g=(\hbdudhat_k^g)^H \mathbf v_k^g=(\hbdudhat_k^g)^H \mathbf J_k^g \hbdudhat_k^g, \ g=1,\cdots, G.
\end{align}



   The main computational complexity for the proposed small-MIMO processing scheme is due to the matrix inversion in  \eqref{eq:vkg}, which is in the order of $\sum_{g=1}^G O(M_g^3)$. Therefore, with appropriate antenna grouping such that $M_g\ll M$, $\forall g$, a significant complexity reduction can be achieved as compared to the full-scale MMSE receiver in \eqref{eq:vkMMSE}, which has the complexity of $O(M^3)$. 
 Next, we show that in the EM-lens enabled system, the full-scale MMSE receiver in \eqref{eq:vkMMSE} reduces to the proposed small-MIMO processing receiver under certain conditions, in which case the computational complexity reduction by the proposed scheme is achieved without any performance loss.

Under Assumption~1, the energy of the incident waves of each UT after passing through the EM lens is focused on a subset of $2\Delta+1$ antennas. We assume that the antennas can be grouped in a way such that for each $\UT_k$, all the antennas with non-zero power belong to the same group, denoted as group $g_k$. 
In this case, the channel covariance matrix $\Rbdud_k$ is then given by a block diagonal structure as $\Rbdud_k= \blkdiag\{\mathbf 0, \Rbdud_{k}^{g_k}, \mathbf 0\}$, where $\Rbdud_k^{g_k}\in \mathbb{C}^{M_{g_k}\times M_{g_k}}$ is the covariance matrix for channels of $\UT_k$ corresponding to the antennas in group $g_k$, and $\mathbf 0$ is an all-zero matrix of appropriate size. Similarly, we have $\hbdudhat_k \hbdudhat_k^H = \blkdiag\{\mathbf 0, \hbdudhat_k^{g_k} (\hbdudhat_k^{g_k})^H, \mathbf 0\}$, and $\Ebdud_k=\blkdiag\{\mathbf 0, \Ebdud_k^{g_k}, \mathbf 0\}$, $\forall k$. By evaluating the MMSE receiver in \eqref{eq:vkMMSE} with the above noted block-diagonal matrices, we can obtain
\begin{align}
\mathbf v_k= \blkdiag\left\{\mathbf J_k^1,\cdots, \mathbf J_k^G\right\}\hbdudhat_k,
\end{align}
and the resulting signal $\hat x_k$ in \eqref{eq:xkhat} becomes
\begin{align}
\hat x_k=\mathbf v_k^H \mathbf y = & \hbdudhat_k^H \blkdiag\left\{\mathbf J_k^1,\cdots, \mathbf J_k^G\right\}\mathbf y\\
=& \sum_{g=1}^G  (\hbdudhat_k^{g})^H\mathbf J_k^g  \mathbf y^g,
\end{align}
which coincides to that obtained by the proposed small-MIMO processing given in \eqref{eq:xkhatsmallMIMO} with $w_k^g=1$, $\forall g$. In other words, under Assumption 1 and with ``ideal'' antenna grouping described above, the proposed low-complexity small-MIMO processing gives the same performance as the full-scale MMSE receiver. In the general scenario where ``ideal'' grouping cannot be attained due to interference coupling across all antennas, simulation results in Section~\ref{sec:SimuSmallMIMO} show that the performance loss due to antenna grouping and intra-group MMSE is marginal with sufficiently separated AoAs of different UTs.

\subsection{Channel Covariance Based Antenna Selection}\label{sec:AS}
While the small-MIMO processing scheme proposed in the previous subsection reduces the computational complexity, it still requires all the $M$ BS antennas to be activated. As $M$ becomes large, it is costly in terms of both hardware implementation and energy consumption to make all antennas operate  at the same time. A practical  low-cost solution is thus  antenna selection (AS) \cite{369}, where the ``best'' subset of $N$ out of $M$ receive antennas are selected for processing the received signals. AS  reduces the number of required RF chains significantly from $M$ to $N$ if $N$ is much smaller than $M$. The optimal AS scheme in general requires instantaneous CSI for all the $M$ antennas, which may be achieved by sequential channel estimation when only $N<M$ RF chains are available \cite{369}. Nevertheless, this would require  an increase of training time by a factor $M/N$ as compared to the case when $M$ RF chains are available, which may significantly compromise the spectral efficiency since less time will be available for data transmission. In this subsection, we propose a new AS scheme that only requires the knowledge of the channel second-order statistics or covariance matrices. As a result, only the channels for the selected antennas need to be estimated instantaneously. 
It turns out that our new AS scheme is particularly suitable for the EM-lens enabled system, in which the channel covariance matrices of UTs with different AoAs vary  significantly due to AoA-dependent energy focusing by the EM lens.

Let $\mathcal M\subset \{1,\cdots, M\}$ denote a subset of the BS antennas, and $\bar \gamma_k^{\mathcal M}$ denote the average SNR lower bound given in \eqref{eq:gammaLB} for $\UT_k$ when only the BS antennas in set $\mathcal M$ are used. From \eqref{eq:gammaLB}, it is evident that $\bar \gamma_k^{\mathcal M}$ depends only on the channel covariance matrices. For a given $\mathcal{M}$, we define the sum rate as $R^{\mathcal M}=\sum_{k=1}^K \log_2(1+\bar \gamma_k^{\mathcal M})$, which gives an approximation for the actual sum rate corresponding to \eqref{eq:Rk} and is used as our performance metric for AS. To find the best $N$ out of $M$ antennas so that $R^{\mathcal M}$ is maximized, an exhaustive search over  $\binom{M}{N}$ number of possible selections is needed, which may incur a high complexity for large $M$ and moderate $N$. We thus propose a low-complexity greedy AS scheme, which is summarized below.

\begin{algorithm}[H]
\caption{Channel Covariance Based Antenna Selection}
\label{A:AS}
\begin{algorithmic}[1]
\STATE  Initialize the set of selected and unselected antennas as $\mathcal S=\emptyset$ and $\mathcal U=\{1,\cdots, M\}$, respectively.
\WHILE{$\mathrm{Card}(\mathcal S)<N$, with $\mathrm{Card}(\cdot)$ denoting the number of elements in a set,}
\STATE Let $n^{\star}=\underset{n\in \mathcal U}{\max} \ R^{\mathcal S \cup \{n\}}$. Update $\mathcal S= \mathcal S \cup \{n^{\star}\}$ and $\mathcal U=\mathcal U \backslash \{ n^{\star}\}$.
\ENDWHILE
\end{algorithmic}
\end{algorithm}

\section{Numerical Results}\label{sec:simulation}
In this section, simulation results are provided to verify our analysis and evaluate our proposed designs in this paper. We consider a single-cell uplink transmission, where the BS is equipped with a $50$-element ULA (i.e., $M=50$) with adjacent antennas separated by $d=\lambda$. The coverage angle of the ULA is set as $\Theta=\pi/3$ so that $\theta_k\in [-\pi/3, \pi/3]$, $\forall k$. The channel vector $\mathbf h_k$  of $\UT_k$ for the system without EM lens is generated based on the CSCG distribution $\mathbf h_k \sim \mathcal {CN}(\mathbf 0,\mathbf R_k)$, where the covariance matrix $\mathbf R_k$  is obtained by the Gaussian PAS as given in \eqref{eq:Gaussian} with the large-scale fading coefficient set as $\beta_k=1$, $\forall k$, and angular spread $\sigma_{\phi}=10^{\circ}$ for all UTs. For the EM-lens enabled system, the spatial power distribution vector $\mathbf a(\theta)$ that varies with the AoA $\theta$ is modeled by \eqref{eq:atheta} and \eqref{eq:fytheta} assuming Gaussian density functions with $\Delta=2$ and $V=0.5d^2$, which corresponds to a power drop by $90\%$ with a distance $3d$ away from the peak power location for a given $\theta$. Moreover, the peak power location is modeled as $\bar y(\theta)=y_{_{\Delta+1}}+\frac{\theta+\pi/3}{2\pi/3}\left(y_{_{ M-\Delta}}-y_{_{\Delta+1}}\right)$, $\theta\in \left[-\pi/3, \pi/3\right]$, so that as the AoA $\theta$ varies from $-\pi/3$ to $\pi/3$, $\bar y(\theta)$ sweeps uniformly between the locations of the $(\Delta+1)$'s and the $(M-\Delta)$'s antenna elements.

\begin{figure}
\centering
\includegraphics[scale=0.35]{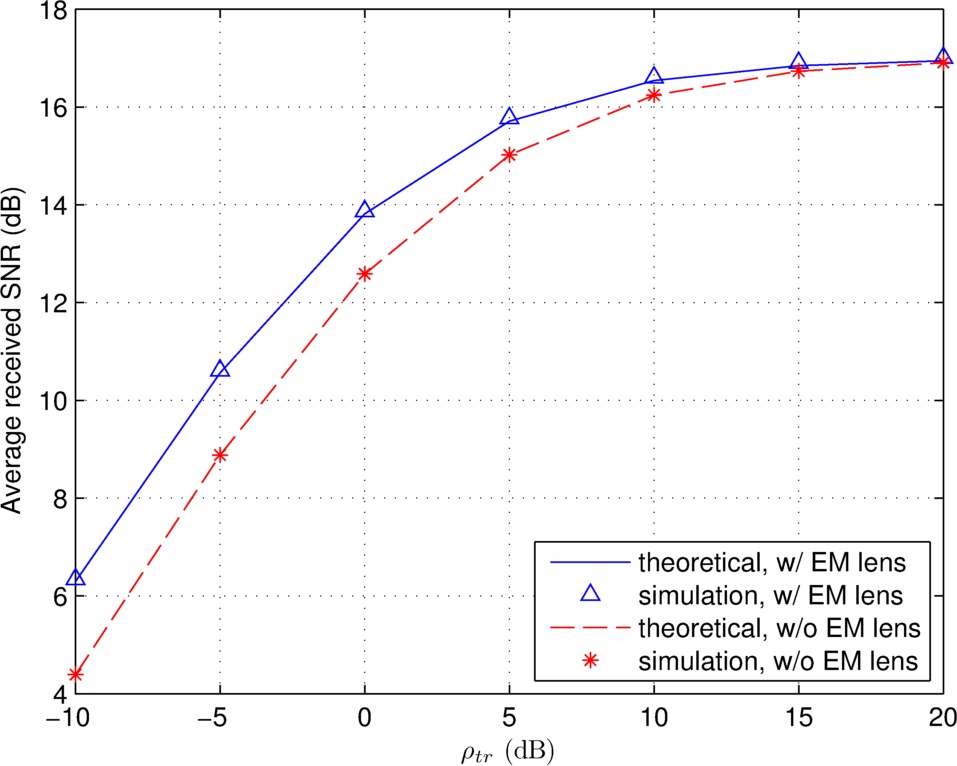}
\caption{Average received SNR versus $\rho_{\ttr}$ in single-user system.}
\label{F:SNRvsRhoSUM50}
\end{figure}


\subsection{Single-User System}
First, we consider a single-user system with nominal AoA $\theta=0$. With the SNR for data transmission set as $\rho_d=0$ dB, Fig.~\ref{F:SNRvsRhoSUM50} plots  the average received SNR  versus the training SNR $\rho_{\ttr}$  for the two systems with versus without EM lens, where the averaging is taken over $10000$ random channel realizations. The theoretical values of $\mathbb{E}[\gamma]$ given in \eqref{eq:fx} are also plotted in the same figure. It is observed that the theoretical and simulation results match perfectly since in single-user system, the average SNR given in \eqref{eq:fx} is exact. It is also observed that the EM-lens enabled system strictly outperforms that without the EM lens at all values of $\rho_{\ttr}$, while the two systems tend to achieve the same average SNR  at sufficiently high $\rho_{\ttr}$, which is in accordance with our analytical result given in Proposition~\ref{prop:rhoInfty}. In addition, it is noted that the performance gain by the EM-lens enabled system is more pronounced when the training power is low, since in this case the energy focusing provided by the EM lens is more beneficial as the limited training power can be concentrated to provide better channel estimation for the most dominant antenna elements.

\begin{figure}
\centering
\includegraphics[scale=0.32]{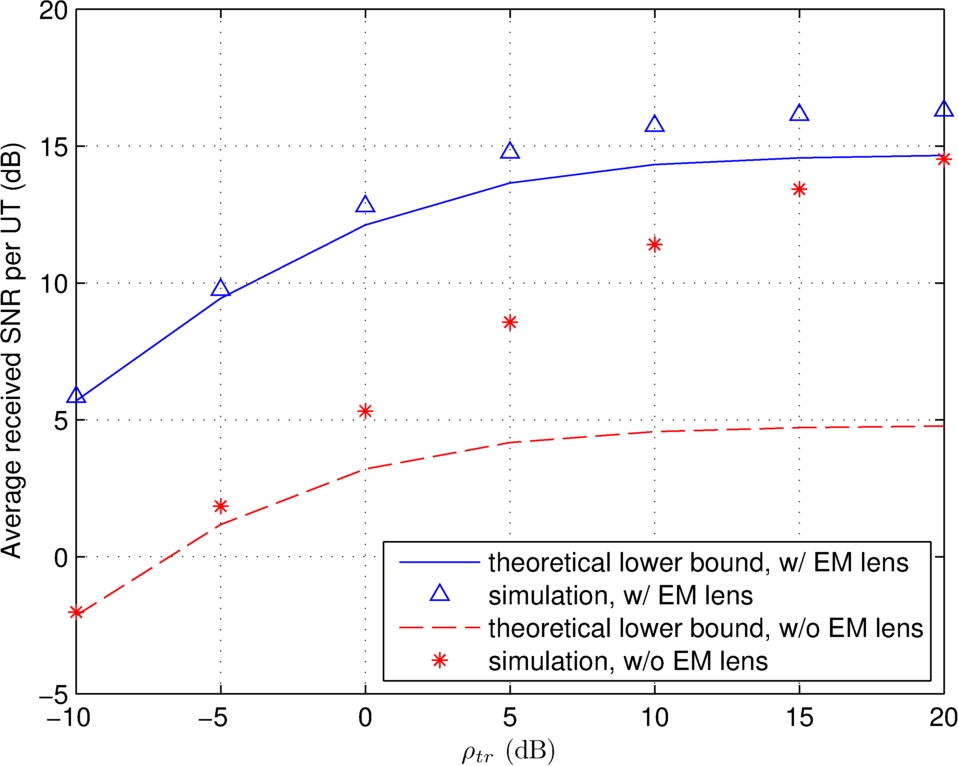}
\caption{Average received SNR versus $\rho_{\ttr}$ in multiuser system with $K=20$ and $M=50$.}
\label{F:SNRvsRhoM50K20}
\end{figure}

\subsection{Multiuser System}\label{sec:MUsystem}
Next, we consider a multiuser system with $K=20$ UTs whose nominal AoAs are equally spaced between $-\pi/3$ and $\pi/3$. With $\rho_d=0$ dB, Fig.~\ref{F:SNRvsRhoM50K20} plots the average received SNR against $\rho_{\ttr}$ for one randomly selected UT. It is observed that  significant performance gains are achieved by the EM-lens enabled system as compared to the conventional system without EM lens. It is interesting to note that, similar to the single-user case, the performance gap for the two systems reduces with the increasing of $\rho_{\ttr}$, which is expected due to the less usefulness  of energy focusing when more training power is available; however, different from the single-user case, the gap in Fig.~\ref{F:SNRvsRhoM50K20} does not diminish to zero even with sufficiently high $\rho_{\ttr}$, which is due to the additional interference-rejection  gain by the EM-lens enabled system in the multiuser case (see Remark~\ref{remark:MURhoInfty}). We have also plotted in Fig.~\ref{F:SNRvsRhoM50K20} the theoretical lower bound $\bar \gamma_k$  given in \eqref{eq:gammaLB}, which is computed solely based on the channel covariance matrices. It is observed that for the EM-lens enabled system, the lower bound $\bar \gamma_k$, which is essentially achieved by simply ignoring the estimated instantaneous channel knowledge of all other UTs,  has a good match with the actual value of  $\mathbb{E}[\gamma_k]$ over all $\rho_{\ttr}$ values. This is expected since thanks  to the spatial interference rejection by the EM lens, ignoring the instantaneous channel knowledge of other users in the MMSE receiver does not harm the performance too much. In contrast, for the system without EM lens,  $\bar \gamma_k$ is significantly lower than $\mathbb{E}[\gamma_k]$, especially at high $\rho_{\ttr}$ regime when the channel estimation is sufficiently accurate. This implies the necessity  of utilizing the instantaneous channel knowledge of all UTs for interference suppression in the system without EM lens due to the more severe interference across all the antennas.

\begin{figure}
\centering
\includegraphics[scale=0.31]{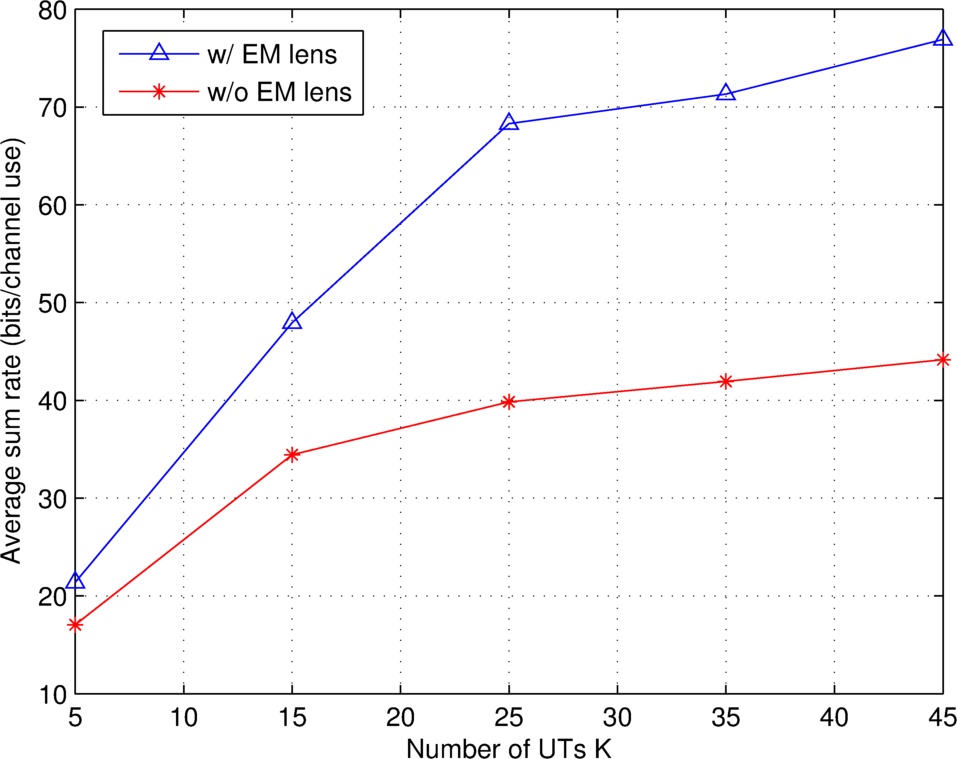}
\caption{Average achievable sum rate versus the number of UTs, $K$.}
\label{F:SumRatevsKM50rhod0rhotr10_UniformAngleSpace}
\end{figure}

In Fig.~\ref{F:SumRatevsKM50rhod0rhotr10_UniformAngleSpace}, the average achievable sum rate versus the number of UTs, $K$, is plotted. For each $K$ value shown in Fig.~\ref{F:SumRatevsKM50rhod0rhotr10_UniformAngleSpace}, $10000$ random channel realizations are simulated, with the nominal AoAs uniformly drawn between $-\pi/3$ and $\pi/3$. The uplink data and training SNRs are set as $\rho_d=0$ dB and $\rho_{\ttr}=10$ dB, respectively. It is observed  that for both systems with and without EM lens, the sum rate increases with the number of UTs $K$, but at a lower speed as $K$ increases. This is expected since when many UTs transmit simultaneously, the sum rate is limited by the inter-user interference and hence further increasing the number of UTs $K$ will not notably improve the sum rate. It is also observed from Fig.~\ref{F:SumRatevsKM50rhod0rhotr10_UniformAngleSpace} that the EM-lens enabled system outperforms the conventional system for any number of UTs, whereas the gain is more pronounced in the regime with larger $K$ values, since spatial interference rejection by the EM lens is more effective when each UT is more severely interfered by other UTs.

\subsection{Small-MIMO Processing}\label{sec:SimuSmallMIMO}
\begin{figure}
\centering
\includegraphics[scale=0.32]{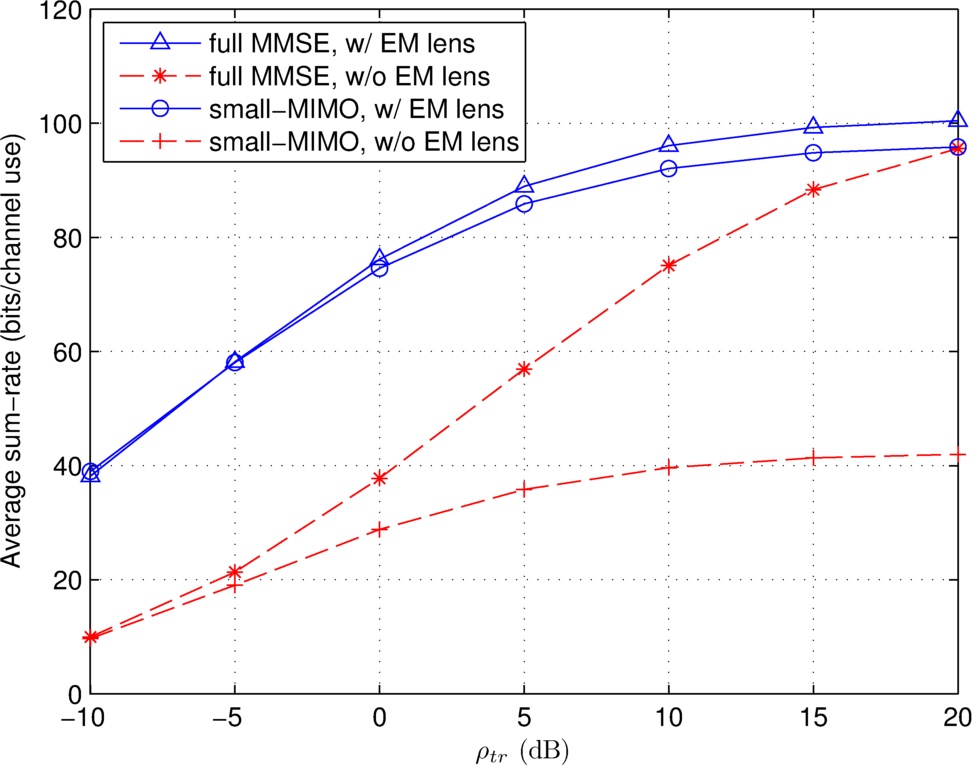}
\caption{Average achievable sum rate by full-scale MMSE versus small-MIMO processing in multiuser system with $K=20$ and $M=50$.}
\label{F:RateM50K20dOneGroupMMSE}
\end{figure}

In this subsection, for a multiuser system with $K=20$ and $M=50$, we provide a performance evaluation for the  small-MIMO processing scheme proposed in Section~\ref{sec:smallMIMO}. The $50$ BS antennas are divided into $10$ groups, each with $5$ elements, i.e., $G=10$, and $M_g=5$, $\forall g$. For each group, the MMSE receiver given in \eqref{eq:vkg} is separately performed and then the output signals for each UT from all the $G$ groups are combined based on MRC. With $\rho_d=0$ dB, Fig.~\ref{F:RateM50K20dOneGroupMMSE} plots the average achievable sum rate versus $\rho_{\ttr}$ for the two systems with and without EM lens. As a benchmark, the performance of the full-scale MMSE receiver given in \eqref{eq:vkMMSE} is also shown in the same figure. It is observed that for the system without EM lens, the proposed small-MIMO processing with intra-group MMSE filtering incurs significant rate loss as compared to the full-scale MMSE processing. In contrast, for the EM-lens enabled system, the performance loss due to antenna grouping and intra-group MMSE is observed to be marginal. 
It is worth pointing out that, with the low-complexity small-MIMO processing, the EM-lens enabled system even outperforms the conventional system without EM lens applied with the full-scale MMSE processing.

\subsection{Antenna Selection}
\begin{figure}
\centering
\includegraphics[scale=0.32]{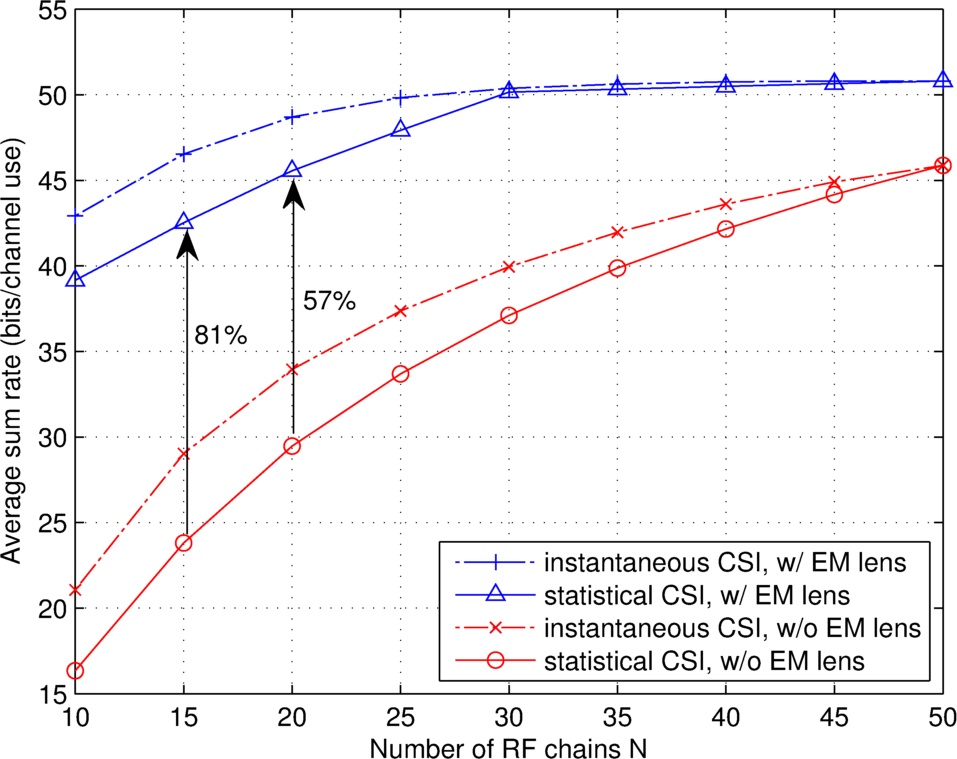}
\caption{Average achievable rate versus the number of RF chains in multiuser system with $K=10$ and $M=50$.}
\label{F:AntSelectM50K10dOne}
\end{figure}

At last, we provide a performance comparison for the two systems with versus without EM lens when AS is applied. We consider the setup with $K=10$ UTs, with their nominal AoAs equally spaced between $-\pi/3$ and $\pi/3$. With $\rho_{\ttr}=10$ dB and $\rho_d=0$ dB, Fig.~\ref{F:AntSelectM50K10dOne} plots the average achievable sum rate  versus the number of active antennas $N$ with the covariance based AS scheme presented in Algorithm~\ref{A:AS}.  As a benchmark, the results for the instantaneous CSI based AS scheme are also plotted. It is observed that for both systems with and without EM lens, the instantaneous CSI based AS achieves higher sum rates than the covariance/statistical CSI based scheme, as expected. However, as discussed in Section~\ref{sec:AS}, to select $N$ out of $M$ antennas, the former scheme generally requires $M/N$ folded more training time in order to obtain the instantaneous CSI for all the $M$ antennas. As a consequence, depending on the channel coherence time in practical systems, the instantaneous CSI based AS may outperform less notably or even perform worse than the statistical CSI based AS when the training overhead is taken into account. For the proposed covariance based AS scheme, it is observed that significant rate gains are achieved by the EM-lens enabled system over that without EM lens. For example, with $N=15$ or $N=20$, a $81\%$ or $57\%$  rate gain is achievable. Moreover, in order to achieve  above $99\%$ of the  maximum rate  in each case,  almost all the $50$ antennas need to be activated for the  system without EM lens, while this number is significantly reduced to $30$ in the case with EM lens, as observed from Fig.~\ref{F:AntSelectM50K10dOne}. It is also observed that for the EM-lens enabled system, activating only $20$ antennas is sufficient to achieve the same sum rate as that of the system without EM lens even when all $50$ antennas are activated.

\section{Conclusion and Future Work}\label{sec:conclusion}
\subsection{Conclusion}
In this paper, we propose a novel antenna
 system design for large-scale MIMO communication, where an EM lens is
 integrated with the large antenna array, termed
EM-lens enabled MIMO. An EM-lens enabled system offers two main benefits, namely energy focusing of the desired signal and  spatial rejection of the interference signal. Under the practical setup with imperfect channel estimation via uplink training, we analytically show the strict improvement on the average received SNR by the EM-lens enabled system. For the single-user case, the performance gain is due to energy focusing of the desired signal, which allows more accurate channel estimation for the most dominant antenna elements. On the other hand, under the multiuser setup, an additional gain due to the spatial interference rejection  is obtained. In order to reduce the signal processing complexity, we propose a new multiuser receiver called small-MIMO processing, which works particularly well for the EM-lens enabled system. Furthermore, when the number of available RF chains is practically less than that  of antennas, a channel covariance based antenna selection scheme is proposed to reduce the hardware and energy costs. Simulation results are presented to validate our analysis and show the great potential  advantages of the EM-lens enabled MIMO system for next generation cellular networks.

\subsection{Future Work}
There are a number of directions along which the developed results in this work can be further investigated, which are briefly discussed as follows.
\begin{itemize}
\item \emph{2D or 3D Array Configurations}: While ULA has been considered in this paper as a proof of concept, the proposed design can be in principle extended to more complicated  2D or 3D array configurations, by applying the corresponding channel models and appropriately characterizing the power distributions for waves passing through the EM lens in the 2D or 3D space. 
\item \emph{Downlink Transmission}:  It is necessary to study the proposed EM-lens enabled system in downlink transmission. For the conventional massive MIMO systems without EM lens, the uplink and downlink channel reciprocity has been widely assumed under time division duplexing (TDD). Since the newly integrated  EM lens to the antenna array is a passive device and thus has a linear and invertible transfer function, the reciprocity property should in principle hold in the EM-lens enabled system. Therefore, the techniques developed in this paper for uplink can be similarly extended to downlink, via exploiting the channel reciprocity.
\item \emph{Millimeter Wave Communication}: Millimeter wave (mmW) communication over the large unused mmW bands between 30 and 300 GHz has its great potential for the next generation wireless communication \cite{486}. To compensate for the severe path loss in mmW communications, large antenna arrays are generally equipped at the transmitter and/or receiver to achieve enormous beamforming gains. The proposed design of the EM-lens enabled MIMO can be applied for mmW communications to further improve the beamforming performance and yet reduce the hardware and signal processing costs. 
\item \emph{User Scheduling}: In order to fully utilize the benefit of the EM-lens enabled system, the BS should assign UTs with sufficiently large AoA separations to the same time/freqeuncy channel  for uplink/downlink transmission. This may require new AoA-based joint space-time and/or space-frequency user scheduling designs for practical systems.
\item \emph{Multi-Cell Systems}: In the multi-cell setup, the ``pilot contamination'' problem is believed to be a key performance limitation factor for massive MIMO systems \cite{470,472,473}. With our proposed EM-lens enabled design, pilot contamination is expected to be effectively mitigated with AoA-aware cooperative user scheduling and pilot assignment by different BSs, thanks to the spatial interference rejection offered by the EM lens. 
\end{itemize}

 \appendices
\section{Overview of Schur-Convex Function and Majorization Theory}\label{A:major}
This appendix provides a very brief overview of Schur-convex functions and majorization theory, on which most analytical results of this paper are based. A more comprehensive treatment of this topic is given by the textbook \cite{457} and its application to signal processing can be found in, e.g., \cite{475}.
\begin{definition}\label{def:majorization}
A vector $\mathbf x \in \mathbb{R}^M$ is said to be majorized by a vector $\mathbf y \in \mathbb{R}^{M}$, denoted as $\mathbf x \prec \mathbf y$, if
\begin{align}
&\sum_{m=1}^N x_{[m]}\leq \sum_{m=1}^N y_{[m]}, \  N=1,\cdots, M-1, \label{eq:major}\\
&\sum_{m=1}^M x_{[m]}= \sum_{m=1}^M y_{[m]},
\end{align}
where $[\cdot]$ is a permutation such that $x_{[1]}\geq x_{[2]}\geq \cdots \geq x_{[M]}$.
\end{definition}

The following is an important though trivial example of majorization.
\begin{lemma}\label{lemma:majortrivial}
For any $\mathbf x\in \mathbb{R}^M$ with $\sum_{m=1}^M x_m=C$, $x_m\geq 0$, $\forall m$, we have
\[
\frac{C}{M}\left[\begin{matrix} 1 & 1 & \cdots & 1\end{matrix}\right]^T \prec \mathbf x \prec \left[\begin{matrix} C & 0 & \cdots & 0\end{matrix}\right]^T.
\]
\end{lemma}

Functions that preserve the preordering of majorization are called \emph{Schur-convex}.
\begin{definition}\label{def:schurCVXFunction}
A real-valued function $g$ defined on a set $\mathcal S \subset \mathbb{R}^M$ is said to be Schur-convex on $\mathcal S$ if
\begin{align}
\mathbf x \prec \mathbf y \text{ on } \mathcal S \Rightarrow g(\mathbf x) \leq g(\mathbf y).
\end{align}
In addition, if $g(\mathbf x)< g(\mathbf y)$ whenever $\mathbf x \prec \mathbf y$ and $\mathbf  y$ is not a permutation of $\mathbf x$, then $g$ is said
to be \emph{strictly Schur-convex} on $\mathcal S$.  
\end{definition}

We then have the following lemma, of which the proof can be found in \cite{457} and thus are omitted here for brevity.
\begin{lemma} \label{lemma:schurcvxTest}
Let $\mathcal{I}\subset \mathcal R$ be  an interval and let $g(\mathbf x)=\sum_{m=1}^M h(x_m)$, where $h: \mathcal I \rightarrow \mathbb R$. If $h$ is (strictly) convex
on $\mathcal I$, then $g$ is (strictly) Schur-convex on $\mathcal I^M$.
\end{lemma}

\begin{lemma}\label{lemma:SchurCVXonD}
Let $\varphi$ be a real-valued continuous function, defined on $\mathcal D$ and continuously differentiable on the interior of $\mathcal D$, where $\mathcal D=\{\mathbf z: z_1\geq \cdots \geq z_M\}$. $\varphi(\mathbf z)$ is then Schur-convex on $\mathcal D$ if and only if $\frac{\partial \varphi(\mathbf z)}{\partial z_m}$ is decreasing in $m=1,\cdots, M$, i.e., the gradient $\nabla \varphi(\mathbf z)\in \mathcal D$, for all $\mathbf z$ in the interior of $\mathcal D$.
\end{lemma}

\begin{lemma}\label{lemma:quadra}
Let $\mathbf S=[s_{mn}]$ be a real symmetric $M\times M$ matrix. Then
\begin{align}
g(\mathbf x)=\mathbf x^T \mathbf S \mathbf x
\end{align}
is Schur-convex on $\mathcal D_+=\left\{\mathbf z: z_1\geq \cdots \geq z_M\geq 0\right\}$ if and only if
\begin{align}
\sum_{n=1}^l(s_{kn}-s_{(k+1)n})\geq 0, \ l=1, \cdots, M, \ k=1, \cdots, M-1. \label{eq:condQuadra}
\end{align}
\end{lemma}

\section{Proof of Proposition~\ref{prop:LOS}}\label{A:LOS}
It can be obtained from \eqref{eq:hm} that in LOS environment where $\theta_{1l}=\theta_1, \forall l$, the covariance matrix $\mathbf R$ for the  single-user channel without EM lens is of rank one and  can be expressed as $\mathbf R=\mathbf b\mathbf b^H$, where $\mathbf b=\sqrt{\beta}\left[\begin{matrix}1 & \cdots & \exp\left(j\frac{2\pi d}{\lambda} (M-1) \sin \theta_1  \right)\end{matrix}\right]^T$. As a result, $\mathbf R$ has only one non-zero
eigenvalue, which is equal to $\tr(\mathbf R)=\beta M$. So is $\Rbdud=\sqrt{\mathbf A}\mathbf R \sqrt{\mathbf A}$ for any $\mathbf A$ satisfying \eqref{eq:trR}, i.e.,
\begin{align}
\boldsymbol \lambda(\Rbdud)=\boldsymbol \lambda(\mathbf R)=\left[\begin{matrix} \beta M & 0 & \cdots & 0 \end{matrix} \right]^T.
\end{align}
Proposition~\ref{prop:LOS} then readily follows from Lemma~\ref{lemma:SchurCVX}.

\section{Proof of Proposition~\ref{prop:lowSNR2}}\label{A:lowSNR2}
To prove Proposition~\ref{prop:lowSNR2}, we first show the following result.
\begin{lemma}\label{lemma:lowSNR}
For the single-user  system with non-LOS channel and $\rho_d+\rho_{\ttr}\ll \frac{1}{\beta M}$,  we have $\bar \gamma(\Rbdud)> \bar \gamma(\mathbf R)$ if the power distribution vector $\mathbf a \neq \mathbf 1$ satisfies
\begin{align}
a_m \geq a_n, \forall  |m-\bar m| \leq |n-\bar m|, \label{eq:condaLowSNR}
\end{align}
where $\bar m\triangleq \left \lceil \frac{M}{2} \right \rceil$ denotes the center of the ULA, which is also the peak power location by assumption.  $\lceil\cdot\rceil$ denotes the ceil operation.
\end{lemma}
\begin{IEEEproof}
Since $\sum_{m=1}^M \lambda_m(\Rbdud)=\beta M$,  we have $\lambda_{\max} (\Rbdud)\leq \beta M$. Therefore, in the low-SNR regime specified in Lemma~\ref{lemma:lowSNR}, we have $\rho_d+\rho_{\ttr}\ll \frac{1}{\beta M} \leq \frac{1}{\lambda_{\max}(\Rbdud)}$; hence \eqref{eq:fx} reduces to $f(\mathbf x)=\rho_d\rho_{\ttr} \sum_{m=1}^M x_m^2$, and the average received SNR can be simplified as
\begin{align}
\bar \gamma(\Rbdud)&=\rho_d\rho_{\ttr} \sum_{m=1}^M \lambda_m^2(\Rbdud)\\
&=\rho_d\rho_{\ttr} \sum_{m=1}^M \lambda_m(\Rbdud^H \Rbdud) \label{eq:eigSq}\\
&=\rho_d\rho_{\ttr} \tr\left(\Rbdud^H \Rbdud\right) \\
&=\rho_d\rho_{\ttr}\sum_{m=1}^M\sum_{n=1}^M \left|[\Rbdud]_{mn} \right|^2\\
&=\rho_d\rho_{\ttr}\sum_{m=1}^M\sum_{n=1}^M a_m a_n |[\mathbf R]_{mn}|^2\\
&=\rho_d\rho_{\ttr}\mathbf a^T \mathbf Q \mathbf a \label{eq:quadra},
\end{align}
where in \eqref{eq:eigSq}, we have used the identity $\lambda(\mathbf X^2)=\lambda^2(\mathbf X)$ and $\Rbdud=\Rbdud^H$; in \eqref{eq:quadra}, the  $M\times M$ real-valued matrix $\mathbf Q$ is defined as $[\mathbf Q]_{mn}=\left|[\mathbf R]_{mn}\right|^2$. Since $\mathbf R$ is the spatial correlation matrix of a ULA, which is Hermitian and Toeplitz $\left(\text{i.e., } [\mathbf R]_{mn}=
 [\mathbf R]_{m+k,n+k}\right)$, it can be shown that $\mathbf Q$ is symmetric and Toeplitz, which can be specified as
 \begin{align}\label{eq:Qmatrix}
 [\mathbf Q]_{mn}=Q_{|m-n|}, \ m,n=1,\cdots, M,
 \end{align}
 where the $M$ numbers that completely determine $\mathbf Q$ satisfy $Q_0\geq Q_1 \geq \cdots \geq Q_{M-1}$.


 In order to prove Lemma~\ref{lemma:lowSNR}, we make use of Lemma~\ref{lemma:quadra}  in Appendix~\ref{A:major} as follows. From \eqref{eq:quadra}, we have
 \begin{align}
 \bar \gamma(\Rbdud)&=\rho_d\rho_{\ttr}\mathbf a^T \mathbf Q \mathbf a =\rho_d\rho_{\ttr} \left(\mathbf \Pi \mathbf a\right)^T\left(\mathbf \Pi \mathbf Q \mathbf \Pi^T \right)\left(\mathbf \Pi \mathbf a\right)\notag \\
 &=\rho_d\rho_{\ttr} \tilde{\mathbf a}^T \tilde{\mathbf Q} \tilde{\mathbf a},
 \end{align}
 where $\mathbf \Pi$ is a permutation matrix such that the elements in the vector $\tilde{\mathbf a}=\mathbf \Pi \mathbf a$ are in non-increasing order, i.e., $\tilde{\mathbf a}\in \mathcal{D}_+$, with the set
 $\mathcal D_+$ defined in Lemma~\ref{lemma:quadra}. With the power distribution vector $\mathbf a$ satisfying the condition given in Lemma~\ref{lemma:lowSNR} and the matrix $\mathbf Q$ given by
 \eqref{eq:Qmatrix}, it can be verified that the resulting matrix $\tilde{\mathbf Q}=\mathbf \Pi \mathbf Q \mathbf \Pi^T$ satisfies the conditions specified in \eqref{eq:condQuadra}. Together with the fact that
 $\mathbf 1 \prec \tilde{\mathbf a}$ on $\mathcal{D}_+$, we have the following result by invoking Lemma~\ref{lemma:quadra}:
 \begin{align}
 \bar \gamma(\mathbf R)=\rho_d\rho_{\ttr}\mathbf 1^T \tilde{\mathbf Q} \mathbf 1  \leq \rho_d\rho_{\ttr} \tilde{\mathbf a}^T \tilde{\mathbf Q} \tilde{\mathbf a}=
  \bar \gamma\left(\Rbdud\right), 
 \end{align}
 where the strict inequality holds when $\mathbf a\neq \mathbf 1$.

 This thus completes the proof of Lemma~\ref{lemma:lowSNR}.
\end{IEEEproof}

Lemma~\ref{lemma:lowSNR} differs from Proposition~\ref{prop:lowSNR2} in that it requires the peak power location occuring at the center of the ULA, which may not be the case when the AoA $\theta \neq 0$. 
Next, we will show that the power distribution vector $\mathbf a$ given in Proposition~\ref{prop:lowSNR2} (or Assumption 1) is essentially equivalent to that in Lemma~\ref{lemma:lowSNR}.


Note that the power distribution vector $\mathbf a$ satisfying Assumption 1 can be written in the form $\mathbf a=\left[\begin{matrix} \mathbf 0_{m^{\star}-1-\Delta}; & \boldsymbol \alpha; & \mathbf 0_{M-m^{\star}-\Delta} \end{matrix} \right]$, where $\boldsymbol \alpha$ has dimension $2\Delta +1$. With \eqref{eq:quadra}, we have
\begin{align}
\bar \gamma(\Rbdud)=\rho_d\rho_{\ttr}\mathbf a^T \mathbf Q \mathbf a = \rho_d\rho_{\ttr} \boldsymbol \alpha^T \mathbf Q' \boldsymbol \alpha,
\end{align}
where $\mathbf Q'\in \mathbb{R}^{(2\Delta+1)\times (2\Delta+1)}$ is obtained from $\mathbf Q$ by deleting the first $m^{\star}-1-\Delta$ rows and columns, as well as the last $M-m^{\star}-\Delta$ rows and columns. With $\mathbf Q$ given by \eqref{eq:Qmatrix}, we have $[\mathbf Q']_{mn}=Q_{|m-n|}$, $1\leq m,n \leq 2\Delta +1$.

 Consider another power distribution vector $\hat {\mathbf a}$ given by $\hat {\mathbf a}=\left[\begin{matrix} \mathbf 0_{\left \lceil (M-2\Delta-1)/2 \right\rceil}; & \boldsymbol \alpha; & \mathbf 0_{\left \lfloor (M-2\Delta-1)/2 \right\rfloor} \end{matrix} \right]$, and the resulting covariance matrix denoted as $\hat{\underline{\mathbf R}}$,  we then have
 \begin{align}
 \bar{\gamma}(\hat{\underline{\mathbf R}})= \rho_d\rho_{\ttr} {\hat {\mathbf a}}^T \mathbf Q \hat {\mathbf a}=\rho_d\rho_{\ttr} \boldsymbol \alpha^T \hat {\mathbf Q} \boldsymbol \alpha,
 \end{align}
 where $\hat {\mathbf Q}$ is obtained from $\mathbf Q$ by deleting the first $\left \lceil (M-2\Delta-1)/2 \right\rceil$ rows and columns, as well as the last $\left \lfloor (M-2\Delta-1)/2 \right\rfloor$ rows and columns with $\lfloor\cdot\rfloor$ denoting the floor operation, which yields $[\hat {\mathbf Q}]_{mn}=Q_{|m-n|}$. Therefore, we have $\mathbf Q'=\hat{\mathbf Q}$, and hence $\bar \gamma(\Rbdud)=\bar{\gamma}(\hat{\underline{\mathbf R}})$. Furthermore, it can be verified that with $\mathbf a$ satisfying the conditions
 specified in Proposition~\ref{prop:lowSNR2}, the newly constructed power vector $\hat{\mathbf a}$ must satisfy the conditions given in Lemma~\ref{lemma:lowSNR}. We thus have
 \begin{align}
 \bar \gamma(\Rbdud)=\bar{\gamma}(\hat{\underline{\mathbf R}})>\bar \gamma(\mathbf R).
 \end{align}

 This thus completes the proof of Proposition~\ref{prop:lowSNR2}.

\section{Proof of Theorem~\ref{theo:gainMU}}\label{A:gainMU}
For a given $\UT_k$ in the EM-lens enabled system, let $\underline{\boldsymbol \xi}\in \mathbb{R}_+^{M}$ denote the average \emph{desired} signal powers received by the $M$ antennas, i.e., the $m$th entry of $\underline{\boldsymbol \xi}$ is given by $\underline{\xi}_m=\beta_k a_m(\theta_k)$, $m=1,\cdots,M$. Also let $\underline{\boldsymbol \kappa}\in \mathbb{R}_+^{M}$ denote the average  \emph{interference} power received, i.e., $\underline{\kappa}_m=\sum_{u\neq k} \beta_u a_m(\theta_u)$, $m=1,\cdots,M$. Similarly, $\boldsymbol \xi$ and $\boldsymbol \kappa$ are defined for the conventional system without EM lens. It is evident that $\boldsymbol \xi=\beta_k \mathbf 1$ and $\boldsymbol \kappa=(\sum_{u\neq k} \beta_u) \mathbf 1$. With Lemma~\ref{lemma:majortrivial} in Appendix~\ref{A:major}, we have
\begin{align}\label{eq:twoMajors}
\boldsymbol \xi \prec \underline{\boldsymbol \xi}, \quad \boldsymbol \kappa \prec \underline{\boldsymbol \kappa}.
\end{align}

Define a function $\psi: \mathbb{R}_+^M \times \mathbb{R}_+^M \rightarrow \mathbb R$ as
\begin{align}
\psi(\mathbf x, \mathbf y)=\sum_{m=1}^M \frac{\rho_{\ttr}\rho_d x_m^2}{x_m \left(\rho_{\ttr}\rho_d y_m+\rho_{\ttr}+\rho_d \right)+\rho_d y_m +1}.
\end{align}
Then the average SNR given in \eqref{eq:gammaMU} for the systems with and without the EM lens are  given by $\bar{\gamma}_k\left(\Rbdud_1,\cdots, \Rbdud_K\right)=\psi(\underline{\boldsymbol \xi}, \underline{\boldsymbol \kappa})$ and  $\bar{\gamma}_k\left(\mathbf R_1,\cdots, \mathbf R_K\right)=\psi({\boldsymbol \xi}, {\boldsymbol \kappa})$, respectively. We need to show that the following two inequalities hold under condition \eqref{eq:condMU}:
\begin{align}
\psi(\boldsymbol \xi, \boldsymbol \kappa)\leq \psi(\underline{\boldsymbol \xi}, \boldsymbol \kappa)\leq \psi(\underline{\boldsymbol \xi}, \underline{\boldsymbol \kappa}).\label{eq:twoIneq}
\end{align}

To show the first inequality in \eqref{eq:twoIneq}, it is noted that since $\boldsymbol \kappa$ has identical entries, the function $\psi(\mathbf x, \boldsymbol \kappa)$ with respect to $\mathbf x$ 
 can be expressed in the form $\psi(\mathbf x, \boldsymbol \kappa)=\sum_{m=1}^M h(x_m)$ for some function $h$. By applying Lemma~\ref{lemma:schurcvxTest} in Appendix~\ref{A:major}, it follows that $\psi(\mathbf x, \boldsymbol \kappa)$ is Schur-convex with respect to $\mathbf x$. Together with \eqref{eq:twoMajors}, the first inequality in \eqref{eq:twoIneq} thus follows.

To show the second inequality in \eqref{eq:twoIneq}, it is noted that with the condition given in \eqref{eq:condMU}, we have $\underline{\xi}_m<\underline{\xi}_n\Rightarrow \underline{\kappa}_m \geq \underline{\kappa}_n$. As a consequence, $\psi(\underline{\boldsymbol \xi}, \underline{\boldsymbol \kappa})$ can be equivalently expressed as
\begin{align}
\psi(\underline{\boldsymbol \xi}, \underline{\boldsymbol \kappa})=\sum_{m=1}^M \frac{\rho_{\ttr}\rho_d \underline{\xi}_{(m)}^2}{\underline{\xi}_{(m)} \left[\rho_{\ttr}\rho_d \underline{\kappa}_{[m]}+\rho_{\ttr}+\rho_d \right]+\rho_d \underline{\kappa}_{[m]} +1},
\end{align}
where $(\cdot)$ is a permutation so that $\underline{\xi}_{(1)}\leq \underline{\xi}_{(2)}\leq \cdots \leq \underline{\xi}_{(M)}$, and $[\cdot]$ is a permutation so that $\underline{\kappa}_{[1]}\geq \underline{\kappa}_{[2]}\geq \cdots \geq \underline{\kappa}_{[M]}$. Define a function $\varphi: \mathcal D\rightarrow \mathbb R$, where $\mathcal D=\{\mathbf z: z_1\geq \cdots \geq z_M\}$, as \begin{align}
\varphi(\tilde {\mathbf y})=\sum_{m=1}^M \frac{\rho_{\ttr}\rho_d \underline{\xi}_{(m)}^2}{\underline{\xi}_{(m)} \left[\rho_{\ttr}\rho_d \tilde y_m +\rho_{\ttr}+\rho_d \right]+\rho_d \tilde{y}_m+1}.
\end{align}
With Lemma~\ref{lemma:SchurCVXonD} in Appendix~\ref{A:major}, it can be verified that $\varphi(\tilde {\mathbf y})$ is a Schur-convex function on $\mathcal D$. Furthermore, we have
\begin{align}
\psi(\underline{\boldsymbol \xi}, \underline{\boldsymbol \kappa})=\varphi(\tilde {\underline{\boldsymbol \kappa}}),
\end{align}
where $\tilde {\underline{\boldsymbol \kappa}}=\boldsymbol \Pi \underline{\boldsymbol \kappa}$ for some permutation $\boldsymbol \Pi$ so that $\tilde {\underline{\boldsymbol \kappa}}\in \mathcal D$. Similarly, we have  $\psi(\underline{\boldsymbol \xi}, {\boldsymbol \kappa})=\varphi(\boldsymbol \kappa)$. With $\boldsymbol \kappa \prec \underline{\boldsymbol \kappa}$ as given in \eqref{eq:twoMajors}, we have $\boldsymbol \kappa\prec \tilde {\underline{\boldsymbol \kappa}}$ on $\mathcal D$ since majorization is invariant to permutation. Together with the Schur-convexity of $\varphi(\tilde {\mathbf y})$ on $\mathcal D$, the second inequality in \eqref{eq:twoIneq} thus follows. 

This thus completes the proof of Theorem~\ref{theo:gainMU}.

\bibliographystyle{IEEEtran}
\bibliography{IEEEabrv,IEEEfull}

\end{document}